\documentclass[pdflatex,sn-basic]{sn-jnl}
\usepackage{titlesec}
\usepackage{natbib}
\usepackage{enumitem}
\usepackage{tabularx}
\usepackage{booktabs}
\usepackage{hyperref}
\usepackage{bookmark}
\usepackage{graphicx}%
\usepackage{multirow}%
\usepackage{amsmath,amssymb,amsfonts}%
\usepackage{amsthm}%
\usepackage{dsfont}
\usepackage{mathtools}
\usepackage{rotating}
\usepackage{comment}
\usepackage{multirow}
\usepackage{geometry}
\usepackage{bm}
\usepackage{mathrsfs}%
\usepackage[title]{appendix}%
\usepackage{xcolor}%
\usepackage{textcomp}%
\usepackage{manyfoot}%
\usepackage{booktabs}%
\usepackage{algorithm}%
\usepackage{algorithmic}
\usepackage{listings}%
\usepackage{bookmark}
\usepackage{anyfontsize}
\usepackage{tikz}
\usepackage{tikz-3dplot}
\usepackage{colortbl} 
\definecolor{lightgray}{gray}{0.9}

\renewenvironment{table}[1][]%
{\tableorg[#1]%
\tablebodyfont%
\renewcommand\footnotetext[2][]{{\removelastskip\vskip3pt%
\let\tablebodyfont\tablefootnotefont%
\hskip0pt\if!##1!\else{\smash{$^{##1}$}}\fi##2\par}}%
}{\endtableorg}

\usepackage{float}

\newtheorem{assumption}{Assumption}
\newtheorem{proposition}{Proposition}
\newtheorem{remark}{Remark}

\begin{document}

\title[Article Title]{Strategic Interactions in Science and Technology Networks: Substitutes or Complements?}

\author*[1]{\fnm{Michael} \sur{Balzer}}\email{michael.balzer@uni-bielefeld.de}

\author*[1]{\fnm{Adhen} \sur{Benlahlou}}\email{adhen.benlahlou@uni-bielefeld.de}

\affil*[1]{\orgdiv{Bielefeld University}, \orgname{Center for Mathematical Economics}, \orgaddress{\street{Universitätsstraße 25}, \city{Bielefeld}, \postcode{33615}, \state{NW}, \country{Germany}}}

\abstract{This paper develops a theory of scientific and technological peer effects to study how individuals' productivity responds to the behavior and network positions of their collaborators across both scientific and inventive activities. Building on a simultaneous equation network framework, the model predicts that productivity in each activity increases in a variation of the Katz–Bonacich centrality that captures within-activity and cross-activity strategic complementarities. To test these predictions, we assemble the universe of cancer-related publications and patents and construct coauthorship and coinventorship networks that jointly map the collaboration structure of researchers active in both spheres. Using an instrumental-variables approach based on predicted link formation from exogenous dyadic characteristics, and incorporating community fixed effects to address endogenous network formation, we show that both authors’ and inventors’ outputs rise with their network centrality, consistent with the theory. Moreover, scientific productivity significantly enhances technological productivity, while technological output does not exert a detectable reciprocal effect on scientific production, highlighting an asymmetric linkage aligned with a science-driven model of innovation. These findings provide the first empirical evidence on the joint dynamics of scientific and inventive peer effects, underscore the micro-foundations of the co-evolution of science and technology, and reveal how collaboration structures can be leveraged to design policies that enhance collective knowledge creation and downstream innovation.}


\maketitle

\section{Introduction} \label{sec:intro}
Innovation is widely recognized as a key driver of economic growth, with research and development (R\&D) playing a central role \citep{romer1990endogenous, aghion1992model}. Science and technology are core components of this process. Scientific research generates new knowledge and increases the efficiency of inventive activities, even when innovations do not emerge directly from scientific breakthroughs \citep{kline2010overview, david1992analysing, nelson1959simple}.

Debates about the relationship between science and technology have evolved substantially. The traditional linear model, which treated science as an exogenous input to technological progress, has gradually given way to a view in which the two domains co-evolve. Empirical work since the mid-1980s documents the strengthening links between scientific research and technological innovation, and this growing interdependence increasingly shapes public policy. Governments have expanded efforts to promote university–industry linkages, often by encouraging technology transfer and reinforcing intellectual property rights for academic discoveries \citep{breschi2010tracing}.

Despite progress in measuring science–technology linkages, existing empirical approaches still overlook the interactions between scientific and inventive activities at the individual level. Understanding how these activities influence each other within and across collaboration networks is essential for capturing the micro-foundations of the co-evolution of science and technological productivities.

This study investigates how interactions with other scientists in a co-authorship network influence individual productivity, as well as how interactions with other inventors in a co-inventorship network affect their inventive output, a dynamic commonly referred to as scientific peer effects. To study these interrelated dynamics, we also examine how productivity in one type of activity relates to productivity in the other at the agent level. A central question in the economics of science and innovation is how an agent’s position within a collaboration network shapes their productivity.

This paper quantifies heterogeneity in peer effects among authors and inventors by studying how network position and individual characteristics jointly shape their productivity.
The microfoundations of this study rest on two fundamental premises. First, agents outcomes can be divided into an idiosyncratic component and a peer effect component. Therefore, a scientist’s productivity is influenced both by their individual characteristics and by the influence of their peers. Second, at the group level, peer effects encompass the collective influence that group members
exert on each other, including the bilateral cross-influences within the group but also across activities. This model formally links an agent’s productivity to their network position captured by a variation of a centrality measure defined
by \citep{bonacich1987power}.

In this paper, we build upon the framework of \cite{chen2018} which establishes the behavioral foundation for the estimation of a simultaneous equation network model \citep{cohen2018}. Within this framework, the main parameter of interest captures the strength of the dyadic influences within the each network but also across these networks. These parameters plays a crucial role in the computation of the Katz-Bonacich centralities, as it serves as the decaying weight for path length. However, estimating this model poses certain challenges.

Identifying peer effects is challenging because it requires accounting for both strategic complementarities and the non-random formation of collaboration ties. At the dyadic level, higher research output is associated with greater effort and stronger collaboration, but agents differ in intrinsic productivity, making heterogeneity essential to capture network synergies. Moreover, collaborations are voluntary and formed around mutual interests; partner selection is therefore endogenous rather than random. An agent’s collaboration decisions reflect strategic expectations about the benefits of working with specific peers, and their network position follows from these choices. As a result, collaboration networks are shaped by agents’ strategic behavior and by unobserved factors that influence both productivity and the propensity to collaborate \citep{mairesse2006measurement, fafchamps2010matching, freeman2014and}. Addressing this complexity presents significant challenges. Prior work has exploited the unexpected deaths of scientists as an exogenous shock to assess the impact of collaboration on productivity
\citep{azoulay2010, azoulay2019does, jaravel2018team, oettl2012reconceptualizing}.

We develop an instrumental variables (IV) approach using a logistic regression to predict link formation probabilities between all pairs of agents. Taking advantage of the panel data structure in the empirical analysis, we introduce communities fixed effects into the social network model to attenuate the potential asymptotic bias caused by the endogenous social adjacency matrix. To further reduce this potential bias, we use the predicted adjacency matrix based on predetermined dyadic characteristics (instead of the observed adjacency matrix) to construct IV for this model. These predictions are based on exogenous dyadic characteristics such as whether the individuals belong to the same institution, share similar demographics, and other relevant characteristics. This approach is robust even in networks where connections are endogenously determined and simplifies the computational complexity by bypassing the need to model the entire network formation process. By focusing on one particular field, the analysis is more tractable, and comparisons between different agents are more valid, as different subfields of the medical science universe may have different research production processes. Our choice thereby mitigates potential confounding across subfields. This allows us to identify the causal effect of collaboration, as well as the impact of simultaneously being an author and an inventor. 

We test the proposed microfoundations using the universe of cancer-related publications and patents, a field where scientific and technological activities are closely connected. Cancer research is highly collaborative and relies heavily on tacit knowledge, which is most effectively transferred through direct interaction \citep{polanyi2009tacit, nonaka2007}. As a result, both publication and patent outputs can be directly linked to collaboration structures \citep{katz1997}. The field is also economically significant: a 1\% annual reduction in mortality for major cancers has been estimated to reduce productivity losses by about \$814 million per year \citep{bradley2008productivity}.

The first result of this study is consistent with the theory, both authors and inventors's productivities are linked to their network position captured by a variation of the Katz-Bonacich centrality. This results highlight the role of strategic complementarities as a driver of productivity in creative activities.

A second central finding shows that higher scientific productivity contributes positively and significantly to an agent’s technological productivity. This reinforces the view that investments in scientific research can yield downstream benefits for technological performance at the agent level, ultimately spilling across the technological network through the strategic complementarities at play.

A third finding shows that higher technological productivity does not contribute significantly to an agent’s scientific productivity. Taken together with the previous result, this pattern aligns with a perspective closer to the linear model, in which scientific activity constitutes a prerequisite to technological advancement, whereas technological output does not appear to exert a significant reciprocal influence on scientific production.

Our paper is related to different strands of the literature. First, from an econometric perspective, recently there has been significant progress in the literature on identification and estimation of social network models (see \cite{bramoulle2020peer}, for a recent survey). To estimate the social-interaction effect, the conventional estimation strategy (see, for example, \cite{bramoulle2009identification, calvo2009peer, liu2010}) uses the exogenous characteristics of the friends’ friends as IV for the behavior of the friends. More closely to our setup, \cite{cohen2018} provide identification conditions based on the intransitivities in the network structure and propose an IV-based estimation strategy exploiting exogenous characteristics of indirect connections. Yet the validity of the standard IVs strategy relies on the assumption that the network structure captured by the adjacency matrix is exogenous. If the adjacency matrix depends on some unobserved variables that are correlated with the error term of the social interaction regression, then the adjacency matrix is endogenous and this IV-based estimator would be inconsistent. Indeed, there is large evidence in literature supporting the endogeneity of social networks (see, for example, \cite{kelejian2014, hsieh2016, graham2017,mele2017,konig2019r, bonhomme2020econometric, lee2021}.

In this paper, we develop a methodology to estimate a model of social interactions that captures multiple types of peer effects. Individuals' choice may be influenced by peers within the same activity (within-activity peer effects), by their own choices in related activities (simultaneity effects), and by peers' choices in related activities (cross-activity peer effects), while also accounting for correlated effects due to shared unobserved environments. We extend the estimator of \cite{cohen2018} to handle potentially endogenous networks and show through Monte Carlo experiments that it accurately recovers the true model parameters allowing us to identify causal interactions across overlapping networks.

Second, by testing the empirical relevance of our model, we contribute to the literature on peer effects in science and innovation. Unlike prior studies that focus on localized coauthorship networks or assume homogeneity \citep{agrawal2017stars, azoulay2010, azoulay2019does, borjas2015peers,  mohnen2022stars, waldinger2010quality, waldinger2012peer}, we examine both direct and peer effects across broader scientific networks and account for individual characteristics. Our analysis show that peer effects extend beyond prominent researchers and vary with factors such as expertise, gender, and past productivity. While firm networks are known to influence innovation outcomes \citep{powell1996interorganizational, ahuja2000collaboration, konig2019r}, individual inventor-level peer effects remain underexplored, at the exception of \cite{zacchia2020knowledge} and \cite{konig2023}. This study is the first to analyze peer effects simultaneously in scientific and inventive activities, highlighting how agents' network positions shape their productivity.

Third, we contribute to the broad literature of science industry interaction. In particular, we depart from the dominant approach that identifies science-industry linkages using patent-publications pairs to study the effects of science on industry and vice versa \citep{arora2021knowledge, arora2023effect,azoulay2019public, li2017applied}. Rather than focusing on this observable paper trail, we study the more direct effects that arise from the interactions among agents who participate in both spheres. 

The rest of the article unfolds as follows. In Section \ref{sec:micro}, we introduce the network game, characterize the equilibrium, and study optimal subsidies policies. Our estimation strategy is described in Section \ref{sec:econ}. Sections \ref{sec:data} and \ref{sec:res} presents the empirical results. Finally, Section \ref{sec:con} concludes.

\section{Analytical framework} \label{sec:micro}
In this section, we first outline the conceptual framework that forms the foundation of the analysis. This outline provides the necessary theoretical background and intuition for the subsequent formal model, which we introduce in the following subsection.

\subsection{Theory}
At the core of the theoretical framework on scientific peer effects is the notion that research productivity is shaped by the collaborative environments in which scientists are involved. Individual output depends not only on talent and effort but also on interactions within networks that facilitate information exchange and the diffusion of ideas. Collaboration enables the transfer of tacit knowledge that is difficult to codify and particularly valuable when research is novel or complex \citep{polanyi2009tacit}. Because such knowledge circulates primarily within close-knit professional communities, proximity to productive peers can generate substantial spillovers in learning and output \citep{ azoulay2010, waldinger2010quality, iaria2018}.

In science, co-authorship is a primary channel through which peer effects influence individual productivity. Collaborative research enables scientists to combine complementary expertise, share resources, and allocate tasks efficiently, often resulting in higher-quality and more influential outputs than individual efforts \citep{azoulay2010, azoulay2019does, borjas2015peers, mohnen2022stars, ductor2014social, oettl2012reconceptualizing, waldinger2010quality, waldinger2012peer, waldinger2016bombs}. Beyond these direct gains, co-authorship facilitates the transfer of tacit knowledge \citep{patel1999expertise}, exposes researchers to new methodologies and problem framings, and embeds them in productive research networks. Consequently, a scientist’s productivity is closely linked to the productivity and expertise of their collaborators.

Peer effects similarly shape inventive activity, as innovation seldom occurs in isolation \citep{wuchty2007, singh2010lone}. Inventors are embedded in social and organizational networks where collaboration through co-invention and team-based R\&D promotes the exchange of both codified and tacit knowledge \citep{singh2005collaborative}. As in science, an inventor’s productivity depends on the productivity, experience, and specialization of their peers. Empirical evidence shows that inventors connected to highly productive collaborators are more likely to generate valuable patents \citep{akcigit2018dancing, balsmeier2023isolating, jaravel2018team, konig2023, poege2025filling}.

Although science and technology often evolve as distinct domains, they are not isolated islands \citep{cockburn1998absorptive,rosenberg1990firms, breschi2010tracing}. They are connected through researchers who participate in both spheres, publishing scientific work while contributing to patented technologies. Their position as both authors and inventors strengthens the system’s absorptive capacity by allowing scientific insights to be translated into technological applications and by feeding practical challenges back into scientific inquiry \citep{cohen1989innovation, fleming2004science}.
Because much scientific knowledge remains tacit, context-dependent, and embedded in specific experimental practices \citep{patel1999expertise}, such bridge individuals play a crucial role in ensuring that knowledge developed in one domain can be effectively understood and used in the other.

In particular, inventors frequently rely on science as a 'map' for problem-solving and idea generation, using scientific insights to guide the development of new technologies and more efficient production methods \citep{cohen2002links, mokyr2011gifts, fleming2004science}. The ability to effectively leverage scientific knowledge is therefore a key determinant of inventive success at the individual, firm, and industry levels \citep{cockburn1998absorptive, zuckerdarby98}.

However, while science is a public good \citep{arrow1962economic}, its use is neither costless nor automatic \citep{rosenberg1990firms}. Inventors face substantial challenges in identifying relevant scientific advances and accessing the tacit knowledge necessary for practical application. Knowledge developed by scientists often remains tacit long after its discovery, and its transfer is costly, making it 'sticky' \citep{von1994sticky} or 'naturally excludable' \citep{zucker1996star}. Although academic scientists codify new knowledge to facilitate dissemination, full codification is not always in their private interest, as it can be time-consuming and costly \citep{nelson1982schumpeterian, zucker2002commercializing}. A prominent example is the development of recombinant DNA, where diffusion was initially slow because much of the relevant knowledge was tacit; mastering the technique required direct interaction with experienced practitioners \citep{zucker1996star}.

However, publication entails economic trade-offs for inventors and their employers. Beyond the competitive risks associated with knowledge disclosure, openness through publication influences the overall intellectual property (IP) strategy. By releasing information into the public domain, inventors reduce their ability to fully appropriate the returns from their discoveries, effectively transferring part of the potential surplus to other \citep{arundel2001relative}. This strategic disclosure can nevertheless generate dynamic benefits by encouraging follow-on innovation, accelerating technology diffusion, and increasing the likelihood of adoption, which may ultimately enhance licensing opportunities and complementary revenue streams \citep{bar2003value}. At the same time, public disclosure can limit the scope for patent protection at the global level, constraining future appropriability of the underlying invention \citep{rotolo2022firms}.

Existing evidence on the impact of inventive efforts on scientific productivity remains mixed, particularly in the context of university–industry collaborations. Such collaborations offer access to funding, specialized resources, and real-world challenges that can stimulate innovation and applied research. Yet, industry’s focus on intellectual property protection, confidentiality, and commercialization can conflict with academic norms of openness and rapid dissemination \citep{merton1973sociology}. As a result, partnerships with industry may delay publication, limit data sharing, and shift incentives away from academic dissemination toward profit-oriented objectives \citep{blumenthal1996relationships,mowery2001growth}.

Empirical findings reflect these competing dynamics. While industry funding can expand the resources available for research, it may also lead to greater secrecy and slower diffusion of results \citep{czarnitzki2015delay,shibayama2012academic}. In some cases, academic output declines as researchers devote more effort to commercialization activities \citep{toole2010commercializing}. Conversely, other studies show that collaboration with industry can enhance productivity by providing access to complementary expertise, new technologies, and applied research questions \citep{goldfarb2008effect,perkmann2009two}.

Collaboration with industrial researchers can further increases research opportunities by providing financial and material resources, proprietary data, and exposure to diverse forms of expertise. These complementarities foster the creation of larger and more heterogeneous research teams. Empirical evidence indicates that academic scientists engaged in industrial collaborations often achieve higher research output and greater visibility within their fields \citep{azoulay2009impact,cockburn1998absorptive}.

Taken together, these mechanisms show that knowledge production and innovation are jointly shaped by the structure of interactions within and across domains. Scientists and inventors operate in interdependent networks in which outcomes depend not only on direct collaboration but also on indirect exposure. 
Co-authorship and co-invention ties govern the circulation of information, skills, and reputation capital, while dual-role researchers facilitate the transfer of both tacit and codified knowledge between science and technology. As a result, productivity dynamics emerge endogenously from the position of individuals within these networks and from the complementarities between scientific and technological environments.

To formalize these ideas, we use a model in which each agent is embedded in two interconnected networks representing their scientific and technological relationships. Within each domain, peer effects arise from local interactions, while cross-domain feedback capture the influence of exposure to activity in the other network. This framework allows us to analyze how individual productivity responds to both local and cross-layer interactions, and how the joint evolution of science and technology depends on the structure of collaboration.

\subsection{Game-theoretical model}
\subsubsection{Network and notation}
Consider two finite sets of agents, $\mathcal{N}_T$ and $\mathcal{N}_S$, representing inventors and scientists, respectively. Their union forms the population $\mathcal{N} = \mathcal{N}_T \cup \mathcal{N}_S = {1, 2, \dots, n}$. The set $\mathcal{N}_T = {1, 2, \dots, n_T}$ denotes inventors, each corresponding to a node in the co-invention network $g_T$. Similarly, $\mathcal{N}_S = {1, 2, \dots, n_S}$ denotes researchers, where each node represents an author in the co-authorship network $g_S$.

Let $\bm{G}_T = (g_{ij})_T$ and $\bm{G}_S = (g_{ij})_S$ denote the $n_T \times n_T$ and $n_S \times n_S$ adjacency matrices describing collaborative ties among inventors and scientists, respectively. By construction, agent $i$ is connected to agent $j$ if and only if $g_{ij}=1$, and $g_{ij}=0$ otherwise. Both matrices are symmetric, binary, and have zero diagonals, reflecting undirected networks without self-collaboration.\\
Finally, we define the cross-domain matrices $\bm{G}_{TS}$ and $\bm{G}_{ST}$, of dimensions $n_T \times n_S$ and $n_S \times n_T$, respectively, which represent links generated by individuals who are both inventors and authors. These matrices indicate which agents span scientific and technological domains, serving as bridges that connect co-authorship and co-invention networks.

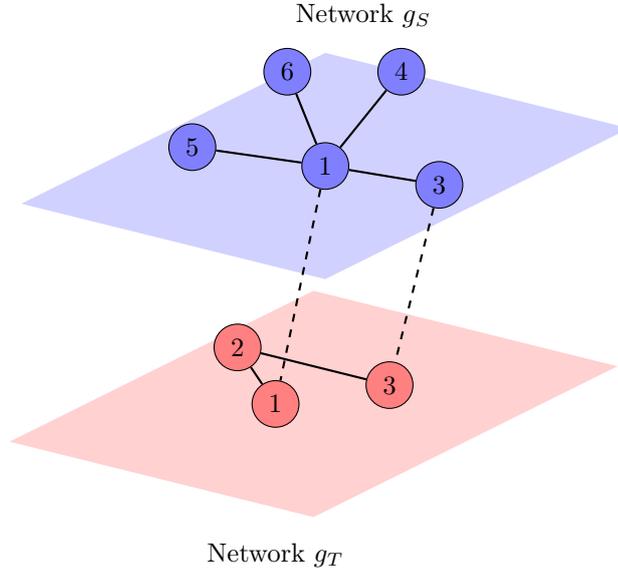
\begin{figure}[!htbp] 
\centering
\begin{tikzpicture}[scale=1]
\fill[blue!30, opacity=0.6] (-4, 1, -1) -- (0, 0, -1) -- (4, 2, -1) -- (0, 3, -1) -- cycle;
\fill[red!30, opacity=0.6] (-3, -1, 2) -- (1, -2, 2) -- (5, 0, 2) -- (1, 1, 2) -- cycle;
\node at (0.5, 3.5, -1) {Network $g_S$};
\node at (0.5, -2.5, 2) {Network $g_T$};
\node (1b) at (0, 1.5, -1) [draw, circle, fill=blue!50, minimum size=5pt] {1};  
\node (3b) at (1.5, 1.25, -1) [draw, circle, fill=blue!50, minimum size=5pt] {3};
\node (4b) at (1, 2.75, -1) [draw, circle, fill=blue!50, minimum size=5pt] {4};
\node (5b) at (-1.75, 1.75, -1) [draw, circle, fill=blue!50, minimum size=5pt] {5};
\node (6b) at (-0.5, 2.75, -1) [draw, circle, fill=blue!50, minimum size=5pt] {6};

\draw[thick] (1b) -- (3b);
\draw[thick] (1b) -- (4b);
\draw[thick] (1b) -- (5b);
\draw[thick] (1b) -- (6b);

\node (1r) at (0.5, -0.5, 2) [draw, circle, fill=red!50, minimum size=5pt] {1};  
\node (2r) at (0, 0.25, 2) [draw, circle, fill=red!50, minimum size=5pt] {2};  
\node (3r) at (2.0, -0.25, 2) [draw, circle, fill=red!50, minimum size=5pt] {3};  

\draw[thick] (1r) -- (2r);
\draw[thick] (2r) -- (3r);

\draw[thick, dashed] (1b) -- (1r);
\draw[thick, dashed] (3b) -- (3r);
\end{tikzpicture}
\caption{Illustrative example of social networks of inventors (red) and researchers (blue) of six agents (nodes) with interrelation. Solid lines represent edges which indicate social connection. Interrelation of both social networks is represented by the dotted lines.} \label{fig:net}
\end{figure}

A basic illustrative example of this model setup can be seen in Figure \ref{fig:net}.

The co-authorship network consists of authors $\mathcal{N}_{s} = \{1,3,4,5,6\}$ connected as shown, while the co-inventor network comprises inventors $\mathcal{N}_{t} = \{1,2,3\}$ with their respective links. Dotted lines indicate individuals active in both scientific and technological production, whereas those without dotted lines participate in only one. The corresponding adjacency matrices are given by
\begin{align*}
&\bm{G}_T = \begin{pmatrix}
0 & 1 & 0 \\
1 & 0 & 1 \\
0 & 1 & 0
\end{pmatrix} & 
&\bm{G}_S = \begin{pmatrix}
0 & 1 & 1 & 1 & 1 \\
1 & 0 & 0 & 0 & 0 \\
1 & 0 & 0 & 0 & 0 \\
1 & 0 & 0 & 0 & 0 \\
1 & 0 & 0 & 0 & 0 
\end{pmatrix} \\
&\bm{G}_{TS} = \begin{pmatrix}
1 & 0 & 0 & 0 & 0  \\
0 & 0 & 0 & 0 & 0 \\
0 & 1 & 0 & 0 & 0 
\end{pmatrix} & 
&\bm{G}_{ST} = \begin{pmatrix}
1 & 0 & 0 \\
0 & 0 & 1 \\
0 & 0 & 0 \\
0 & 0 & 0 \\
0 & 0 & 0 
\end{pmatrix}.
\end{align*}
Let $\bm{y}_{i} = \big(y_{i}^{(T)}, y_{i}^{(S)}\big)$ denote the joint production level of agent $i$. If an agent does not participate in a given network, the corresponding production level is zero by default. The vector
\begin{equation*}
\bm{y}_{-i} = \big(\bm{y}_{1}, \dots, \bm{y}_{i-1}, \bm{y}_{i+1}, \dots, \bm{y}_{n}\big)   
\end{equation*}
represents the technology and science production levels chosen by all agents other than $i$.

\subsubsection{Utility}
Given the network structure, agents decide on their levels of production in scientific and technological activities, with preferences captured by the following utility function, based on \cite{chen2018}
\begin{align}\label{eq:uti}
    u_{i}(\bm{y}_{i}, \bm{y}_{-i}) &= y_{i}^{(T)} \alpha_{i}^{(T)} + y_{i}^{(S)} \alpha_{i}^{(S)} - \left\{\frac{1}{2}\left(y_{i}^{(T)}\right)^{2} + \frac{1}{2}\left(y_{i}^{(S)}\right)^{2} + \beta y_{i}^{(S)} y_{i}^{(T)}\right\} \notag \\
    &\quad + \lambda_{T}\sum_{j = 1}^{n_T}(g_{ij})_{T} y_{i}^{(T)}y_{j}^{(T)} + \lambda_{S}\sum_{j = 1}^{n_S}(g_{ij})_{S} y_{i}^{(S)}y_{j}^{(S)}  \notag \\
     &= \underbrace{y_{i}^{(T)}\alpha_{i}^{(T)} - \frac{1}{2}\left(y_{i}^{(T)}\right)^{2} + \lambda_{T}\sum_{j = 1}^{n_T}(g_{ij})_{T} y_{i}^{(T)}y_{j}^{(T)}}_{\text{net proceeds from technology}}  \\ 
    &+ \underbrace{y_{i}^{(S)}\alpha_{i}^{(S)} - \frac{1}{2}\left(y_{i}^{(S)}\right)^{2}  + \lambda_{S}\sum_{j = 1}^{n_S}(g_{ij})_{S} y_{i}^{(S)}y_{j}^{(S)} }_{\text{net proceeds from science}} \notag \\
    &- \underbrace{\beta y_{i}^{(S)} y_{i}^{(T)}}_{\text{magnitude of interdependence}} \notag.
\end{align}

Here, $\alpha_i^{(S)}$ and $\alpha_i^{(T)}$ represent the agent’s intrinsic productivity in science and technology, respectively, and $\beta \in (-1; 1)$ indicates the complementarity or substitutability between these activities. The terms $\lambda_T \in \mathbb{R^+}$ and $\lambda_S \in \mathbb{R^+}$ capture the intensity of network effects, representing the influence of other agents' efforts on an agent’s own productivity.

Previous work has adopted similar utility functions to study scientific and technological activities in isolation. In the scientific domain, \cite{BAscience} shows that authors’ productivity is driven by Katz–Bonacich centrality, which captures strategic complementarities and outperforms traditional centrality measures. This finding suggests that peer effects arise primarily from effort complementarities rather than from knowledge diffusion. In the technological domain, \cite{benlahloutech} applies the same framework to inventor networks and finds that productivity is likewise shaped by strategic complementarity, rather than by knowledge flows alone, providing empirical support for the relevance of strategic complementarities within each domain.

The within network  peer effect component exhibits heterogeneity that arises endogenously from the positions of individuals within each network, even if they are ex ante identical. The second partial derivatives of the utility function with respect to the production  $y_i^{(S)}$  (respectively $y_i^{(T)}$) and $y_j^{(S)}$ (respectively $y_j^{(T)}$) are given by

\begin{equation*}
 \frac{\partial^2 u_i}{\partial y_i^T \partial y_j^T} = \lambda_T (g_{ij})_T \geq 0, \quad \frac{\partial^2 u_i}{\partial y_i^S \partial y_j^S} = \lambda_S (g_{ij})_S \geq 0.
\end{equation*}
When agent $i$ and $j$ collaborate, the positive value of the cross derivative $\lambda_S$ (respectively $\lambda_T$) signifies strategic complementarity in production. Conversely, if agent $i$ and $j$ do not collaborate, the cross derivative is zero. In our context, a positive $\lambda_S$ (respectively $\lambda_T$) indicates that when agents $i$ and $j$ collaborate, that is, $g_{ij}^{(S)} = 1$ or $g_{ij}^{(T)} = 1$, an increase in $j$'s production leads to a rise in  $i$'s marginal utility from production.

The term $\beta y_i^{(S)} y_i^{(T)}$ accounts for the interconnection between science and technology efforts. Particularly, the second derivative of the utility function with respect to both production levels
\begin{equation*}
 \frac{\partial^{2}u_i}{\partial y_{i}^{(T)} y_{i}^{(S)}} = -\beta,
\end{equation*}
shows that technology and science activities are substitutes if $\beta$ is positive. In contrast, a negative $\beta$ indicates that the activities are complements. Consequently, both activities are independent if and only if $\beta = 0$ holds.

\subsubsection{The peer effect game}
Let's now describe the Nash equilibrium of the game, where each agent $i=1$ selects simultaneously their own production level $y_{i} = (y_i^{s} \geq 0, y_i^{t} \geq 0)$. In equilibrium, each agent maximizes their utility, and their best-response function are given by:
\begin{align}
    y_i^{(T)} &= \alpha_i^{(T)} - \beta y_i^{(S)} + \lambda_T \sum_{j=1}^n \left(g_{ij}\right)_T y_j^{(T)} \label{eq:rp1} \\
    y_i^{(S)} &= \alpha_i^{(S)} - \beta y_i^{(T)} + \lambda_S \sum_{j=1}^n \left(g_{ij}\right)_S y_j^{(S)} \label{eq:rp2}.
\end{align}
Here, the agent’s productivity comprises two distinct effects: two network-specific effect and an idiosyncratic one. In other words, individual behavior can break down into two components: an exogenous part and an endogenous peer effect component that depends on the agent under consideration.

Let $\bm{y}^{(T)}_* = \left(y_{1*}^{(T)}, \cdots, y_{n_T*}^{(T)} \right)^{'}$, $\bm{y}^{(S)}_* = \left(y_{1*}^{(S)}, \dots, y_{n_S*}^{(S)} \right)^{'}$, $\bm{\alpha^{(T)}} = \left(\alpha_{1}^{(T)}, \dots, \alpha_{n_T}^{(T)} \right)^{'}$ and $\bm{\alpha^{(S)}} = \left(\alpha_{1}^{(S)}, \dots, \alpha_{n_S}^{(S)} \right)^{'}$ denote the compact vector form of joint production levels in equilibrium and ability for each agent. Furthermore, let $\bm{I}_{n_T}$ be the $n_T\times n_T$ and $\bm{I}_{n_S}$ the $n_S\times n_S$ identity matrices corresponding to the size of the social networks. To derive the unique Nash equilibrium, denote by $v(\cdot)$ the eigenvalues of any matrix. Let Assumption \ref{ass:net} hold.

\begin{assumption} \label{ass:net}
    $\max\big(\lambda_{T}v(\bm{G}_T), \lambda_{S}v(\bm{G}_S)\big) < 1 - |\beta|$.\
\end{assumption}

\begin{proposition}[\cite{chen2018}] \label{pro:chen}
Suppose that Assumption \ref{ass:net} holds. Then, for any $\bm{\alpha}^{S}$ and $\bm{\alpha}^{T}$, a unique Nash equilibrium exists given by
\begin{align}
    \begin{bmatrix}
    \bm{y}^{(T)}_* \\
    \bm{y}^{(S)}_*
    \end{bmatrix} = \begin{bmatrix}
    \bm{I}_{n_T} - \lambda_{T} \bm{G}_{T} &  \beta \bm{G}_{TS} \\
    \beta \bm{G}_{ST} & \bm{I}_{n_S} - \lambda_{S} \bm{G}_{S}
    \end{bmatrix}^{-1} \begin{bmatrix}
    \bm{\alpha}^{(T)} \\
    \bm{\alpha}^{(S)}
    \end{bmatrix}.
\end{align}
\end{proposition} 

The Assumption \ref{ass:net} used in Proposition \ref{pro:chen} implies that the impact of network complementarities must be sufficiently small relative to the individual convexity of costs. This condition prevents positive feedback loops triggered by such complementarities from escalating without bound. Note that this condition doesn’t directly constrain the absolute values of these cross-effects, but rather their relative magnitude. Network complementarities are assessed through the compound index
$\max\big(\lambda_{T}v(\bm{G}_T), \lambda_{S}v(\bm{G}_S)\big)$, where $\lambda_{S}$ (respectively $\lambda_{T}$) represents the intensity of each non-zero cross-effect, while $v(\bm{G}_T)$ (respectively $v(\bm{G}_S)$) captures the overall pattern of these positive cross-effects. Proposition \ref{pro:chen} establishes a connection between an author’s equilibrium production and their network position, as quantified by their Katz-Bonacich centrality.

\begin{remark}
  We show in the Appendix \ref{app:plan} that in a game where the planner chooses the effort levels, in order to maximise the sum of utilities the equilibrium efforts levels are higher. In particular, their are computed with a doubled peer effects parameters ($2\lambda_{T}$ and $2\lambda_{T}$).  
\end{remark}
\section{Empirical strategy} \label{sec:econ}
This section extends the econometric network model of \cite{cohen2018} by proposing an adjusted version that accounts for endogenous interactions, heteroscedastic errors, and interaction effects across two distinct networks. To this end, a so-called interrelated simultaneous equations network model is proposed where identification threats are identified. Afterwards, two-stage least squares (2SLS) and endogeneity corrected two-stage least squares (2SLS-EC) estimators are proposed. 

\subsection{The interrelated simultaneous equations network model}
Let the observed data be partitioned into $\bar{c}_{T}$ communities in technology $c_{T} = \{1, \cdots, \bar{c}_{T}\}$, each comprised of $n_{T,c_T}$ inventors which are characterized by a network with social adjacency matrix $\bm{G}_{T,c_T} = \{\left(g_{ij, c_T}\right)_T\}$. Similarly, the $\bar{c}_{S}$ communities in science $c_{S} = \{1, \cdots, \bar{c}_{S}\}$ are each comprised of $n_{S,c_S}$ inventors and characterized by a network with social adjacency matrix $\bm{G}_{S,c_S} = \{\left(g_{ij, c_S}\right)_S\}$. Moreover, assume that agents idiosyncratic ability in each community $\alpha_{i, c_T}$ and $\alpha_{i, c_S}$ is a linear function of their observables characteristics given by 
\begin{align*}
\alpha^{(T)}_{i,c_T} &= \mu^{(T)}_{c_T} + \bm{x}^{(T)\prime}_{i,c_T} \bm{\gamma}_T + \epsilon^{(T)}_{i,c_T} \\
\alpha^{(S)}_{i,c_S} &= \mu^{(S)}_{c_S} + \bm{x}^{(S)\prime}_{i,c_S} \bm{\gamma}_S + \epsilon^{(S)}_{i,c_S}
\end{align*}
where $\mu^{(T)}_{c_T}$, $\mu^{(S)}_{c_S}$ are the network fixed effects, $\bm{x}^{(T)\prime}_{i,c_T}$, $\bm{x}^{(S)\prime}_{i,c_S}$ exogenous variables, $\gamma_T$, $\gamma_S$ the corresponding coefficients and $\bm{\epsilon}_{i,c_T}$ as well as $\bm{\epsilon}_{i,c_S}$ the idiosyncratic random innovations unique to agents. The specification of the econometric network model follows the equilibrium best reply functions of the network game given by Equations \ref{eq:rp1} and \ref{eq:rp2} introduced in Section \ref{sec:micro}. Thus, the econometric network model has a clear microfoundation and is given in each community by
\begin{align}
    y_{i, c_T}^{(T)} &=  \lambda_T \sum_{j=1}^{n_{T,c_T}} \left(g_{ij, c_T}\right)_T y_{j, c_T}^{(T)} + \lambda_{TS} y_{i}^{(S)} +  \bm{x}^{(T)\prime}_{i,c_T} \bm{\gamma}_T + \mu^{(T)}_{c_T} + \epsilon^{(T)}_{i,c_T} \label{eq:best_stats1} \\
    y_{i, c_S}^{(S)} &=  \lambda_S \sum_{j=1}^{n_{S,c_S}} \left(g_{ij, c_S}\right)_S y_{j, c_S}^{(S)} + \lambda_{ST} y_{i}^{(T)} + \bm{x}^{(S)\prime}_{i,c_S} \bm{\gamma}_S + \mu^{(S)}_{c_S} + \epsilon^{(S)}_{i,c_S} \label{eq:best_stats2}
\end{align}
where $y_{i, c_T}^{(T)}$, $y_{i, c_S}^{(S)}$ are the outputs of agents and $\lambda_{ST}$, $\lambda_{TS}$ the substitutability or complementarity in the respective science and technology activities. Note that the parameter $\beta$ occurring in Equation \ref{eq:uti} is replaced in the econometric model by the parameters $\lambda_{TS}$ and $\lambda_{ST}$.

Let $\bm{y}_{c_T}^{(T)} = \left(y_{1, c_T}^{(T)}, \cdots, y_{n_{T,c_T}, c_T}^{(T)}\right)^\prime$, $\bm{y}_{c_S}^{(S)} = \left(y_{1, c_S}^{(S)}, \cdots, y_{n_{S,c_S}, c_S}^{(S)}\right)^\prime$, $\bm{X}^{(T)}_{c_T} = \left(\bm{x}^{(T)\prime}_{1,c_T}, \cdots, \bm{x}^{(T)\prime}_{n_{T,c_T},c_T} \right)^\prime$, $\bm{X}^{(S)}_{c_S} = \left(\bm{x}^{(S)\prime}_{1,c_S}, \cdots, \bm{x}^{(S)\prime}_{n_{S,c_S},c_S} \right)^\prime$, $\bm{\epsilon}^{(T)}_{c_T} = \left(\epsilon^{(T)}_{1,c_T}, \cdots, \epsilon^{(T)}_{n_{T,c_T},c_T}\right)^\prime$ and $\bm{\epsilon}^{(S)}_{c_S} = \left(\epsilon^{(S)}_{1,c_S}, \cdots, \epsilon^{(S)}_{n_{S,c_S},c_S}\right)^\prime$. Furthermore, assume that $\bm{\iota}_{n_{T,c_T}}$, $\bm{\iota}_{n_{S,c_S}}$ are vectors of ones and $\bm{G}_{TS, c_S}$, $\bm{G}_{ST, c_T}$ the social network interaction matrices. Additionally, $\text{diag}\{A_{l}\}$ denotes a block diagonal matrix in which the diagonal blocks are $n_l\times m_l$ matrices $A_{l}$. Hence, for a data set with $\bar{c}_S$ and $\bar{c}_T$ communities, let $\bm{y}^{(T)} = \left(\bm{y}_{1}^{(T)'}, \dots, \bm{y}_{c_T}^{(T)'} \right)'$, $\bm{y}^{(S)} = \left(\bm{y}_{1}^{(S)'}, \dots, \bm{y}_{c_S}^{(S)'} \right)'$, $\bm{\mu}_{T} = \left(\mu^{(T)}_{1}, \dots,\mu^{(T)}_{c_T}\right)$, $\bm{\mu}_{S} = \left(\mu^{(S)}_{1}, \dots, \mu^{(S)}_{c_S} \right)$,
$\bm{X}_{T} = \left(\bm{X}^{(T)}_{1}, \dots, \bm{X}^{(T)}_{c_T} \right)^\prime$, 
$\bm{X}_{S} = \left(\bm{X}^{(S)}_{1}, \dots, \bm{X}^{(S)}_{c_S} \right)^\prime$, 
$\bm{\epsilon}_T = \left(\bm{\epsilon}^{(T)\prime}_{1}, \dots, \bm{\epsilon}^{(T)\prime}_{c_T}\right)^\prime$, $\bm{\epsilon}_S = \left(\bm{\epsilon}^{(S)\prime}_{1}, \dots, \bm{\epsilon}^{(S)\prime}_{c_S}\right)^\prime$, $\bm{G}_T = \text{diag}\{\bm{G}_{T,c_T}\}_{c_T = 1}^{\bar{c}_T}$, $\bm{G}_S = \text{diag}\{\bm{G}_{S,c_S}\}_{c_S = 1}^{\bar{c}_S}$, $\bm{G}_{TS} = \text{diag}\{\bm{G}_{TS,c_S}\}_{c_S = 1}^{\bar{c}_S}$, $\bm{G}_{ST} = \text{diag}\{\bm{G}_{ST,c_T}\}_{c_T = 1}^{\bar{c}_T}$, $\bm{L}_T = \text{diag}\{\bm{\iota}_{n_{T,c_T}}\}_{c_T = 1}^{\bar{c}_T}$ and $\bm{L}_S = \text{diag}\{\bm{\iota}_{n_{S,c_S}}\}_{c_S = 1}^{\bar{c}_S}$. Then, the econometric network model given by Equations \ref{eq:best_stats1} and \ref{eq:best_stats2} can be written in vector-matrix form as
\begin{align}
    \bm{y}^{(T)}  &= \lambda_T \bm{G}_T \bm{y}^{(T)} + \lambda_{TS} \bm{G}_{TS} \bm{y}^{(S)}  +  \bm{\mu}_{T} \bm{L}_T + \bm{\gamma}_T \bm{X}_{T} + \bm{\epsilon}_T  \\
    \bm{y}^{(S)} &= \lambda_S \bm{G}_{S} \bm{y}^{(S)} + \lambda_{ST} \bm{G}_{ST} \bm{y}^{(T)} + \bm{\mu}_{S} \bm{L}_S + \bm{\gamma}_S \bm{X}_{S} + \bm{\epsilon}_S.
\end{align}

The network-fixed effects $\bm{\mu}_{T}$ and $\bm{\mu}_{S}$ are allowed to depend on $\bm{G}_{T}$ (respectively $\bm{G}_{S}$), $\bm{X}_{T}$ (respectively $\bm{X}_{S}$) and $\bm{G}_{TS}$ (respectively $\bm{G}_{ST}$) as in a fixed effect panel data model. In cases where the number of communities $\bar{c}_T$, $\bar{c}_S$ is large, the incidental parameter problem may arise \citep{neyman1948}. To eliminate this problem, network fixed effects are removed through the within-transformations $\bm{J}_{c_T} = \bm{I}_{n_{T,c_T}} - \tfrac{1}{n_{T,c_T}} \bm{\iota}_{n_{T,c_T}} \bm{\iota}_{n_{T,c_T}}'$ and $\bm{J}_{c_S} = \bm{I}_{n_{S,c_S}} - \tfrac{1}{n_{S,c_S}} \bm{\iota}_{n_{S,c_S}} \bm{\iota}_{n_{S,c_S}}'$ such that $\bm{J}_T = \text{diag}\{\bm{J}_{c_T}\}_{c_T = 1}^{\bar{c}_T}$
and $\bm{J}_S = \text{diag}\{\bm{J}_{c_S}\}_{c_S = 1}^{\bar{c}_S}$. The econometric network model is then expressed as
\begin{align}
    \bm{J}_T\bm{y}^{(T)}  &= \lambda_T \bm{J}_T \bm{G}_T \bm{y}^{(T)} + \lambda_{TS} \bm{J}_T \bm{G}_{TS} \bm{y}^{(S)}  + \bm{\gamma}_T \bm{J}_T \bm{X}_{T} + \bm{J}_T \bm{\epsilon}_T  \label{eq:mod1} \\
    \bm{J}_S \bm{y}^{(S)} &= \lambda_S \bm{J}_S \bm{G}_{S} \bm{y}^{(S)} + \lambda_{ST} \bm{J}_S \bm{G}_{ST} \bm{y}^{(T)} + \bm{\gamma}_S \bm{J}_S \bm{X}_{S} + \bm{J}_S \bm{\epsilon}_S \label{eq:mod2}. 
\end{align}
Assessing the effects of collaborations ($\lambda_{T}$ and $\lambda_{S}$) and inter-network effect ($\lambda_{ST}$ and $\lambda_{TS}$) is often fraught with econometric challenges. Thus, identification threats and the proposed estimation strategy are based on the transformed model given by Equation \ref{eq:mod1} and \ref{eq:mod2}.


\subsection{Identification threats}
The estimation of parameters in the econometric network model faces several challenges, as highlighted by \cite{manski1993identification}. Peer effects are particularly difficult to identify due to reflection bias, selection bias, and unobserved correlated effects \citep{brock2001discrete, moffitt2001policy}, which create challenges for estimating the parameters in Equations \ref{eq:mod1} and \ref{eq:mod2}.

First, simultaneity arises from feedback loops between agents. An agent’s actions can influence their peers, and these reciprocal effects create simultaneity bias. Second, selection bias can result from non-random sorting of agents into communities based on observable or unobservable characteristics. Third, endogenous collaboration occurs when links form based on anticipated returns rather than randomly, potentially confounding the estimation of peer effects ($\lambda_{T}$ and $\lambda_{S}$).

Correlated effects or common shocks present an additional challenge. Agents within the same community may behave similarly due to shared environmental or community-level factors, which can be mistaken for peer effects. To address this, it is standard to include community fixed effects, which capture unobserved factors influencing behavior. Conceptually, these fixed effects are consistent with a two-step link formation model: first, individuals self-select into communities based on community-specific characteristics; second, links form within communities based on observable agent traits. Community fixed effects therefore partially mitigate bias arising from sorting.

The game-theoretical model emphasizes that collaboration structure itself shapes productivity. Collaborations are often formed based on expectations about outcomes, making the peer network endogenously determined. If factors influencing both collaboration and productivity are omitted, elements of the adjacency matrices may correlate with the error term. For instance, highly curious agents may cluster together, and curiosity can influence both collaboration and outcomes, creating unobserved confounding.

These challenges correspond to \cite{manski1993identification}’s notion of correlated effects, where similarities among co-inventors reflect pre-existing characteristics rather than genuine peer influence. Endogenous co-inventorship ties introduce additional bias that may distort estimation. Addressing these issues requires careful model specification and an appropriate estimation strategy.
\subsection{Estimation}
Given the network econometric model by Equations \ref{eq:mod1} and \ref{eq:mod2}, our aim is to reliably estimate $\bm{\delta}_T = \left(\lambda_T, \lambda_{TS}, \bm{\gamma}_T^{'}\right)^{'}$ and $\bm{\delta}_S = \left(\lambda_S, \lambda_{ST}, \bm{\gamma}_S^{'}\right)^{'}$. The econometric network model can be written more compactly as 
\begin{align}
    \bm{y}^{(T)}  &=  \bm{Z}_T  \bm{\delta}_T +  \bm{\epsilon}_T \\
    \bm{y}^{(S)} &=  \bm{Z}_{S} \bm{\delta}_S +  \bm{\epsilon}_S 
\end{align}
where $\bm{Z}_T = \left[\bm{G}_{T} \bm{y}^{(T)}, \bm{G}_{TS} \bm{y}^{(S)}, \bm{X}_T\right]$ and $\bm{Z}_S = \left[\bm{G}_{S} \bm{y}^{(S)}, \bm{G}_{ST} \bm{y}^{(T)}, \bm{X}_S\right]$. The transformed econometric network model is given in as similar way as
\begin{align}
    \bm{\tilde{y}}^{(T)}  &= \bm{\tilde{Z}}_T  \bm{\delta}_T + \bm{\tilde{\epsilon}}_T \\
    \bm{\tilde{y}}^{(S)} &= \bm{\tilde{Z}}_S \bm{\delta}_S + \bm{\tilde{\epsilon}}_S 
\end{align}
where $\bm{\tilde{y}}^{(T)} = \bm{J}_T \bm{y}^{(T)}$, $\bm{\tilde{Z}}_T = \bm{J}_T \bm{Z}_T$, $\bm{\tilde{\epsilon}}_T = \bm{J}_T \bm{\epsilon}_T$, $\bm{\tilde{y}}^{(S)} = \bm{J}_S \bm{y}^{(S)}$, $\bm{\tilde{Z}}_S = \bm{J}_S \bm{Z}_S$ and $\bm{\tilde{\epsilon}}_S = \bm{J}_S \bm{\epsilon}_S$. 
The structural parameters and coefficients in $\bm{\delta}_T$ and  $\bm{\delta}_S$ can be estimated via maximum likelihood, 2SLS or generalized method of moments estimators \citep{wang2018, guo2019, drukker2023, egger2023}.

For 2SLS, estimation is based on linear moments conditions $\bm{\tilde{H}}_T \bm{\tilde{\epsilon}}_T(\bm{\delta}_T)$ and $\bm{\tilde{H}}_S \bm{\tilde{\epsilon}}_S(\bm{\delta}_S)$ where $\bm{\tilde{H}}_T = \bm{J}_T \bm{H}_T$ and $\bm{\tilde{H}}_S = \bm{J}_S \bm{H}_S$ are transformed matrices of IV. If social adjacency matrices are exogenous, some intransitives in the networks exist and full rank conditions of the matrices of IV hold, then the structural parameters and coefficients are identified. However, since $\bm{\tilde{H}}_T$ and $\bm{\tilde{H}}_S$ involve unknown parameters, the IV matrices are infeasible. Following \cite{liu2014, baltagi2015, cohen2018} and \cite{guo2019}, approximate $n_T \times K$, $n_S \times K$ matrices of feasible IV with exogenous social adjacency matrices can be obtained via a linear combination by applying Neumann series approximation on the reduced form of the econometric network model and are given as $\bm{\tilde{H}}_{T,K}$ and $\bm{\tilde{H}}_{S,K}$ with the approximation error diminishing quickly as $K$ increases. Afterwards, let $\bm{P}_{\bm{\tilde{H}}_{T,K}} = \bm{\tilde{H}}_{T,K} \left(\bm{\tilde{H}}_{T,K}^{'}\bm{\tilde{H}}_{T,K}\right)^{-1}\bm{\tilde{H}}_{T,K}^{'}$ and $\bm{P}_{\bm{\tilde{H}}_{S,K}} = \bm{\tilde{H}}_{S,K} \left(\bm{\tilde{H}}_{S,K}^{'}\bm{\tilde{H}}_{S,K}\right)^{-1}\bm{\tilde{H}}_{S,K}^{'}$ be the projection matrices, then corresponding 2SLS estimators are obtained as
\begin{align}
\bm{\hat{\delta}}_{T,\text{2SLS}} &= \left(\bm{\tilde{Z}}_T^{\prime} \bm{P}_{\bm{\tilde{H}}_{T,K}}  \bm{\tilde{Z}}_T\right)^{-1} \bm{\tilde{Z}}_T^{\prime} \bm{P}_{\bm{\tilde{H}}_{T,K}}  \bm{\tilde{y}}^{(T)} \\
\bm{\hat{\delta}}_{S,\text{2SLS}} &= \left(\bm{\tilde{Z}}_S^{\prime} \bm{P}_{\bm{\tilde{H}}_{S,K}}  \bm{\tilde{Z}}_S\right)^{-1} \bm{\tilde{Z}}_S^{\prime} \bm{P}_{\bm{\tilde{H}}_{S,K}} \bm{\tilde{y}}^{(S)}.
\end{align}

However, if there exists an unobserved individual factor affecting both the productivity and the collaboration between agents, then social adjacency matrices become endogenous, and the feasible IV matrices $ \bm{\tilde{H}}_{T,K}$ and $\bm{\tilde{H}}_{S,K}$ are no longer valid. For instance, think of the possibility that high-ability authors may choose high-potential papers to work on. From working on high-potential papers, they also have a better chance to meet other high ability co-authors. Similarly, one can think about the possibility that inventors choose their co-inventors based on a set of skills that are useful to tackle certain specific issue. As a result, estimating the structural parameters in Equations \ref{eq:mod1} and \ref{eq:mod2} without handling endogeneity of the social adjacency matrices $\bm{G}_{T}$ and $\bm{G}_{S}$ would suffer from a self-selection bias. To resolve the self-selection bias, we propose to use predicted probabilities of link formation based on exogenous variables which replace observed social adjacency matrices  in the 2SLS estimation procedure \citep{kelejian2014, hsieh2016, graham2017, mele2017, lee2021}.

Assuming the existence of agent-specific characteristics that influence link formation, their effects on the conditional probability of forming a link can be incorporated into a latent-utility model for each community $c_T$ and $c_S$:
\begin{align}
\left(g_{ij, c_T}\right)_T &= \mathds{1} \left\{\tau_{0}^{(T)} + \tau_{1}^{(T)} w_{ij,c_T}^{(T)} + u_{i,c_T}^{(T)} + u_{j,c_T}^{(T)} + \kappa_{ij,c_T}^{(T)} > 0 \right\} \\
\left(g_{ij, c_S}\right)_S &= \mathds{1} \left\{\tau_{0}^{(S)} + \tau_{1}^{(S)} w_{ij,c_S}^{(S)} + u_{i,c_S}^{(S)} + u_{j,c_S}^{(S)} + \kappa_{ij,c_S}^{(S)} > 0 \right\} 
\end{align}
where $w_{ij,c_T}^{(T)}$, $w_{ij,c_S}^{(S)}$ are the deterministic components of the latent-utility depending on observable characteristics of both agents, $\tau_{0}^{(T)}$, $\tau_{1}^{(T)}$, $\tau_{0}^{(S)}$ and $\tau_{1}^{(S)}$ the corresponding coefficients,  $u_{j,c_T}^{(T)}$, $u_{j,c_S}^{(S)}$ unobserved heterogeneity and $\kappa_{ij,c_T}^{(T)}$ and $\kappa_{ij,c_S}^{(S)}$ the stochastic component of the utility. Assuming $\kappa_{ij,c_T}^{(T)}$, $\kappa_{ij,c_S}^{(S)}$ follow Gumbel distribution, the link formation probabilities are obtained as
\begin{align}
\mathbb{P}\left(\left(g_{ij, c_T}\right)_T = 1\right) &= \frac{\exp\left(\tau_{0}^{(T)} + \tau_{1}^{(T)} w_{ij,c_T}^{(T)} + u_{i,c_T}^{(T)} + u_{j,c_T}^{(T)}\right)}{1 + \exp\left(\tau_{0}^{(T)} + \tau_{1}^{(T)} w_{ij,c_T}^{(T)} + u_{i,c_T}^{(T)} + u_{j,c_T}^{(T)}\right)} \\
\mathbb{P}\left(\left(g_{ij, c_S}\right)_S  = 1\right) &= \frac{\exp\left(\tau_{0}^{(S)} + \tau_{1}^{(S)} w_{ij,c_S}^{(S)} + u_{i,c_S}^{(S)} + u_{j,c_S}^{(S)}\right)}{1 + \exp\left(\tau_{0}^{(S)} + \tau_{1}^{(S)} w_{ij,c_S}^{(S)} + u_{i,c_S}^{(S)} + u_{j,c_S}^{(S)}\right)}.
\end{align}
Network endogeneity arises if the unobserved individual heterogeneity terms of the agents in the network formation are correlated with the idiosyncratic random innovations of the network econometric model. $\bm{G}_T$ and $\bm{G}_S$ are then endogenous, linear moment conditions are violated and feasible IV matrices are no longer valid.

One way of resolving this issue is to explicitly omit unobserved heterogeneity from the network formation model which is the root of the selection bias. Particularly, the approach proposed by \cite{lee2021} relies on estimating the coefficients in the network formation via a logistic regression  
\begin{align}
\left(\hat{g}_{ij, c_T}\right)_T &= \frac{\exp\left(\hat{\tau}_{0}^{(T)} + \hat{\tau}_{1}^{(T)} w_{ij,c_T}^{(T)}\right)}{1 + \exp\left(\hat{\tau}_{0}^{(T)} + \hat{\tau}_{1}^{(T)} w_{ij,c_T}^{(T)}\right)} \\
\left(\hat{g}_{ij, c_S}\right)_S  &= \frac{\exp\left(\hat{\tau}_{0}^{(S)} + \hat{\tau}_{1}^{(S)} w_{ij,c_S}^{(S)}\right)}{1 + \exp\left(\hat{\tau}_{0}^{(S)} + \hat{\tau}_{1}^{(S)} w_{ij,c_S}^{(S)}\right)}.
\end{align}
where $\hat{\tau}_{0}^{(T)} $, $\hat{\tau}_{1}^{(T)}$, $\hat{\tau}_{0}^{(S)}$, $\hat{\tau}_{1}^{(S)}$ are the estimated coefficients and $\left(\hat{g}_{ij, c_T}\right)_T$ as well as $\left(\hat{g}_{ij, c_S}\right)_S$ are the predicted probabilities of link formation. Since $\bm{\hat{G}}_{T,c_T} = \{\left(\hat{g}_{ij, c_T}\right)_T\}$, $\bm{\hat{G}}_{S,c_S} = \{\left(\hat{g}_{ij, c_S}\right)_S\}$ and hence $\bm{\hat{G}}_T = \text{diag}\{\bm{\hat{G}}_{T,c_T}\}_{c_T = 1}^{\bar{c}_T}$, $\bm{\hat{G}}_S = \text{diag}\{\bm{\hat{G}}_{S,c_S}\}_{c_S = 1}^{\bar{c}_S}$, the consequences of network endogeneity in the feasible IV matrices can be mitigated by replacing social adjacency matrices $\bm{G}_T$ and $\bm{G}_S$ occurring in the IV matrices by the predicted counterparts $\bm{\hat{G}}_T$ and $\bm{\hat{G}}_S$ \citep{kelejian2014, lee2021}. 

To ensure that the predicted adjacency matrices are uniformly bounded in absolute values, the following normalization are applied $\frac{\bm{\hat{G}}_T}{\hat{d}_T}$ and $\frac{\bm{\hat{G}}_S}{\hat{d}_S}$ where $\hat{d}_T = \max \left\{ \max_{i=1} \sum_{j=1}^{n_T} \left(\hat{g}_{ij}\right)_T, \; \max_{j=1} \sum_{i=1}^{n_T} \left(\hat{g}_{ij}\right)_T \right\}$, $\hat{d}_S = \max \left\{ \max_{i=1} \sum_{j=1}^{n_S} \left(\hat{g}_{ij}\right)_S, \; \max_{j=1}^{n_S} \sum_{i=1}^{n_S} \left(\hat{g}_{ij}\right)_S \right\}$ and the social adjacency matrices in the IV matrices are replaced by the predicted counterparts yielding $\hat{\tilde{\bm{H}}}_{T,K}$ and $\hat{\tilde{\bm{H}}}_{S,K}$. Hence, the 2SLS-EC estimators are given by 
\begin{align}
\bm{\hat{\delta}}_{T,\text{2SLS-EC}} &= \left(\bm{\tilde{Z}}_T^{\prime} \bm{P}_{\hat{\tilde{\bm{H}}}_{T,K}} \bm{\tilde{Z}}_T\right)^{-1} \bm{\tilde{Z}}_T^{\prime} \bm{P}_{\hat{\tilde{\bm{H}}}_{T,K}} \bm{\tilde{y}}^{(T)} \\
\bm{\hat{\delta}}_{S,\text{2SLS-EC}} &= \left(\bm{\tilde{Z}}_S^{\prime} \bm{P}_{\hat{\tilde{\bm{H}}}_{S,K}} \bm{\tilde{Z}}_S\right)^{-1} \bm{\tilde{Z}}_S^{\prime} \bm{P}_{\hat{\tilde{\bm{H}}}_{S,K}} \bm{\tilde{y}}^{(S)}.
\end{align}
It is worth to note that the consistency of the proposed estimator does not rely on the consistency of the estimates of the logistic regression. Details and discussions on the assumptions of the econometric network model, the calculation process of the matrices of IV as well as the asymptotic properties can be found in Appendix \ref{app:2sls}.

\section{Data and network definition} \label{sec:data}
Our analysis focuses on authors and inventors in cancer research, a domain where collaboration is prevalent and innovation is both measurable and consequential. The production of knowledge in this field relies heavily on tacit expertise and direct interaction, making collaborations essential for the generation and transfer of new ideas \citep{nonaka2007, polanyi2009tacit}. Scientific output can be assessed through publications, where both quantity and quality provide meaningful indicators of productivity \citep{katz1997}. Similarly, innovation in cancer-related technologies is extensively patented, as sectors such as pharmaceuticals, chemicals, and medical devices exhibit high propensities to patent—over two thirds of innovations are formally protected—compared with less than 15\% in industries such as food or textiles \citep{cohen2003protecting}. Consequently, patent records serve as a reliable proxy for inventive activity and also accurately reflect the collaborative structure of innovation, as only contributors with substantial inventive input are listed as coinventors. This dual focus on publications and patents allows us to comprehensively capture patterns of collaboration and productivity in a field characterized by intensive scientific and technological interaction.

\subsection{Data sources}
To construct the coauthorship network in science $\bm{G}_{S}$, we use data from the MEDLINE publication database. The analysis focuses exclusively on research articles, excluding non-research content such as book reviews, editorials, and letters. Managed by the U.S. National Library of Medicine, MEDLINE is the leading bibliographic database for the life sciences and includes nearly all journal articles published since 1946. The database provides detailed metadata for each publication, including author names, publication dates, journal titles, grant acknowledgments, and Medical Subject Headings (MeSH) terms. MeSH terms constitute a controlled vocabulary assigned by professional indexers specializing in specific biomedical domains. This indexing system ensures a standardized and content-based classification of research topics, providing an objective representation of each publication’s scientific scope. The procedure for identifying cancer-related publications is described in Appendix \ref{app:cancer}.

To construct the coinventorship network in technology $\bm{G}_{T}$, we use the Cancer Moonshot Patent Data \citep{frumkin2016} compiled from the United States Patent and Trademark Office (USPTO). This dataset contains detailed information on published patents related to cancer research and development between 1976 and 2016, covering approximately 270,000 patent documents. The patents span a broad range of technological domains, including pharmaceuticals, diagnostics, surgical instruments, data analysis tools, and genome-based inventions.

Accurate individual identification across patents and publications is essential for constructing inventor and author collaboration networks. Personal names are not unique identifiers and often vary across records due to spelling differences, common names, or incomplete affiliation data. In \textsc{Medline}, for example, nearly two-thirds of authors have ambiguous names, with many sharing the same last name and first initial \citep{smalheiser2009}. This "John Smith" problem has motivated the development of disambiguation algorithms that identify individuals based on name and contextual information. As summarized in \citet{kang2009}, author name disambiguation remains a persistent challenge in bibliometrics.

To address name ambiguity in publication data, we rely on the approach developed by \citet{xu2020building}, who constructed the PubMed Knowledge Graph (PKG) to improve author identification. Their algorithm clusters articles to infer authorship using probabilistic similarity measures across coauthors, topics, and affiliations, and has demonstrated high accuracy, particularly for NIH-funded scientists. The method extends the probabilistic model introduced by \citet{torvik2005probabilistic}, which assumes that articles authored by the same individual share more observable attributes than those authored by different individuals. Building on this framework, \citet{li2014disambiguation} developed a corresponding inventor name disambiguation system for United States patented innovations. A key contribution of their work is the linkage between disambiguated author and inventor identifiers, which enables the construction of cross-domain adjacency matrices $\bm{G}_{ST}$ and $\bm{G}_{TS}$.

To complement the publication and patent datasets, we incorporate metadata compiled by the Torvik research group to construct demographic and institutional variables at both the author and inventor levels.\\
Ethnicity is inferred using the Ethnea tool, which classifies individuals. In principle, the Ethnea tool from the University of Illinois determines authors’ ethnicity based on their names.\footnote{It is available at \url{http://abel.lis.illinois.edu/cgi-bin/ethnea/search.py.}.} Unlike ancestry-based algorithms, \textit{Ethnea} emphasizes contemporary nationality and allows for dual ethnicities, reflecting the increasing prevalence of mixed backgrounds resulting from migration, marriage, and cultural assimilation.\\
Gender is assigned using the \textit{Genni} tool, which employs a probabilistic model that links first names to gender while conditioning on the ethnicity inferred from the surname. For example, the first name \textit{Andrea} is classified as female when associated with a French surname but as male when linked to an Italian surname.\\
Author and inventor affiliations are derived from the \textit{MapAffil} dataset, which links affiliation strings in PubMed records to standardized institutional names, cities, and geocodes worldwide. The version used in this study is based on a 2016 snapshot of PubMed and covers approximately 63\% of all affiliation instances prior to 2015. For inventors, we further identify employers using ownership reassignment data from the USPTO’s patent assignment database. Prior to September 2012, inventors were legally designated as patent owners, but as rights are typically assigned to employers by contract, these transfers are recorded as formal reassignments \citep{marco2015uspto}. For inventors who also appear as scientific authors, we cross-reference MapAffil data to harmonize institutional affiliations across patents and publications.

\subsection{Variables definition}
\subsubsection{Author productivity} We measure individual productivity for both authors and inventors using value-weighted indicators that account for heterogeneity in the significance of scientific and technological outputs. For authors, productivity is defined as the number of publications weighted by the number of citations each paper receives, further adjusted by the impact factor of the citing journal to capture the quality of scientific recognition. For inventors, productivity is measured as the number of patents per inventor, weighted by a breakthrough index that proxies for technological significance. Following \citet{kelly2021measuring}, the breakthrough index is derived through textual analysis of patent documents to construct a high-dimensional measure of innovation. This metric predicts future patent citations and correlates with patent market value as estimated by \citet{kogan2017technological}, while not being restricted to patents granted to publicly traded firms. These adjustments are necessary since neither publications nor patents are of equal value, and simple output counts fail to capture differences in scientific or technological impact.

\subsubsection{IV Construction}
To measure homophily among collaborators in science and technology, we build on the multidimensional proximity framework proposed by \citet{boschma2005proximity}. We assess proximity across four dimensions: cognitive, social, institutional, and cultural, excluding geographical proximity due to its strong correlation with institutional proximity. For both coauthors and coinventors, these measures capture latent forms of similarity that may influence collaboration choices.

\textbf{Cognitive proximity.} Cognitive proximity reflects the similarity of individuals’ knowledge bases and research interests, which shape the likelihood of collaboration \citep{breschi2003knowledge}. For authors, we infer research interests using Medical Subject Headings (MeSH) terms from cancer-related publications produced during the first three years of their careers and compute pairwise cosine similarity. Due to data constraints and the resulting drastic reduction in sample size, we do not compute cognitive proximity for inventors.

\textbf{Social proximity.} Social proximity captures indirect social connections that facilitate trust and reduce uncertainty about potential collaborators’ abilities \citep{uzzi1996sources, breschi2009mobility, fafchamps2010matching}. To mitigate endogeneity concerns, we exclude direct past collaborations and instead rely on indirect ties, such as shared coauthors or coinventors, observed during the five years preceding the beginning of the period.

\textbf{Institutional proximity.} Institutional proximity captures whether two individuals are affiliated with the same organization, reflecting opportunities for face-to-face interaction and informal knowledge transfer \citep{dahlander2013ties, long2014patterns}. We identify institutional affiliations using the \textit{MapAffil} dataset, harmonized across both scientific and patent records.

\textbf{Cultural proximity.} Cultural proximity reflects shared norms, values, and linguistic or social backgrounds that may facilitate communication and collaboration \citep{dahlander2013ties, freeman2015collaborating}. We proxy cultural proximity using ethnicity as inferred from the \textit{Ethnea} tool, which assigns contemporary ethnic categories based on individuals’ names.

\subsection{Network definition}
To prepare the data for estimation, we implement a systematic sample selection procedure. First, the analysis is restricted to patents filed within three-year periods to ensure temporal comparability across observations. Second, we exclude continuation patents, which may artificially inflate measures of inventive productivity by extending the scope or duration of earlier inventions. Third, we limit the sample to patents classified under one of the following categories in the Cancer Moonshot database: \textit{Drugs and Chemistry}, \textit{Model Systems and Animals}, \textit{Cells and Enzymes}, and \textit{DNA, RNA, or Protein Sequences}. Fourth, we restrict the sample to inventors and authors with complete demographic and affiliation information, specifically ethnicity, gender, and institutional affiliation, and with nonzero patenting or publication outcomes. Fifth, for scientists, we restrict the analysis to research and non-clinical publications to ensure consistency in the measurement of scientific output. Finally, we require at least three years of prior research activity to construct MeSH-based vectors of research interests and a positive publication output over the considered period.

Our analysis focuses on authors and inventors active in cancer research, a domain characterized by intensive collaboration and measurable research output through both publications and patents. Collaboration is central to scientific and technological progress, as it facilitates the exchange of knowledge and the recombination of ideas across disciplines and institutions. Much of this knowledge remains tacit and localized within tightly connected research and development teams, making collaborative networks a critical channel for innovation and discovery.


To identify tightly knit groups within the largest component, we employ the modularity-based community detection algorithm developed by \citet{traag2019louvain}, which partitions the network so as to approximately minimize the modularity criterion. The resulting communities are highly clustered, with balanced partitions comprising several groups of comparable size. For the coauthorship network, we construct multiple networks using different values of the resolution parameter, which are subsequently used in the econometric analysis to verify that our results are robust to the choice of community structure. We restrict the analysis to communities containing at least 15 authors to ensure sufficient within-group variation. All selected communities are then pooled to form the coauthorship network under study.

An analogous procedure is applied to the coinventorship network, where two inventors are linked if they are listed as co-inventors on at least one cancer-related patent. We again apply the modularity-based algorithm of \citet{traag2019louvain} to partition the giant component and retain only communities with at least 15 inventors. For inventor communities, the partitions are not sensitive to the resolution parameter, so we use a single value of the parameter in the analysis. These inventor communities are then pooled to form the coinventorship network under study.

We construct the innovation and science networks over non-overlapping time windows, assigning earlier periods to patenting and subsequent periods to publishing. This reflects empirical evidence that research projects leading to patentable inventions often produce related scientific publications with a delay. In many cases, these outputs are developed concurrently but disclosed sequentially, as firms typically impose confidentiality requirements on their academic collaborators to protect intellectual property. Prior studies document that industry-linked research is systematically associated with higher secrecy, reduced knowledge sharing, and substantial publication delays \citep{lee2000sustainability, thursby2002selling, blumenthal1996relationships, blumenthal1997withholding, louis2001entrepreneurship, evans2010industry, shibayama2012academic, czarnitzki2015delay}. For instance, \cite{blumenthal1996relationships} report that 82\% of life-science firms supporting academic research required temporary confidentiality to allow for patent applications, and more than half extended it beyond the filing period. Similarly, \cite{czarnitzki2015delay} find that industry sponsorship increases the likelihood of publication delay from 14\% to 33\% and of secrecy from 11\% to 35\%. These patterns suggest that scientific articles linked to a given set of inventions tend to appear later, justifying our use of lagged publication windows relative to innovation periods. The majority of agents either in science or technology in the sample are male, which aligns with the gender composition typically observed in STEM fields. The
average seniority in either science or industry network is quite high, indicating that most agents are relatively experienced.

\begin{table}[!htbp] \centering 
\caption{Descriptive statistices for the Technology network over the period 2006-2009} 
\label{} 
\begin{tabular}{@{\extracolsep{5pt}}lccccc} 
\\[-1.8ex]\hline 
\hline \\[-1.8ex] 
Variable & \multicolumn{1}{c}{N} & \multicolumn{1}{c}{Mean} & \multicolumn{1}{c}{St. Dev.} & \multicolumn{1}{c}{Min} & \multicolumn{1}{c}{Max} \\ 
\hline \\[-1.8ex] 
Asinh(Weighted production) & 947 & 1.75 & 0.72 & 0.69 & 4.52 \\ 
Asinh(Past citations weighted) & 947 & 4.48 & 2.97 & 0.00 & 11.59 \\ 
Male & 947 & 0.77 & 0.42 & 0 & 1 \\ 
Patenting experience & 947 & 6.48 & 6.76 & $-$3 & 33 \\ 
\hline \\[-1.8ex] 
\end{tabular} 
\end{table}

\begin{table}[!htbp]
\centering
\caption{Descriptive statistics for the Science network 2-year period (2010--2011)}
\label{table:statistiques}
\resizebox{\textwidth}{!}{%
\begin{tabular}{|c|c|c|c|c|c|c|}
\hline
\textbf{Resolution Parameter} & \textbf{Variable} & \textbf{Mean} & \textbf{S.D.} & \textbf{Min} & \textbf{Max} & \textbf{N} \\
\hline

$\phi = 4$ 
& Asinh(Production)              & 5.595 & 1.453 & 1.323 & 10.199 & 1,236 \\
& Asinh(Hit Non clinical)        & 5.124 & 2.333 & 0     & 9.990  & 1,236 \\
& Male                           & 0.753 & 0.431 & 0     & 1      & 1,236 \\
& Asinh(Experience in the field) & 2.727 & 1.197 & 0     & 4.820  & 1,236 \\
\hline

$\phi = 4.5$
& Asinh(Production)              & 5.619 & 1.419 & 1.397 & 9.640  & 1,111 \\
& Asinh(Hit Non clinical)        & 5.113 & 2.328 & 0     & 11.064 & 1,111 \\
& Male                           & 0.757 & 0.429 & 0     & 1      & 1,111 \\
& Asinh(Experience in the field) & 2.700 & 1.205 & 0     & 4.820  & 1,111 \\
\hline

$\phi = 5$
& Asinh(Production)              & 5.605 & 1.425 & 1.397 & 10.199 & 1,203 \\
& Asinh(Hit Non clinical)        & 5.111 & 2.325 & 0     & 9.668  & 1,203 \\
& Male                           & 0.762 & 0.426 & 0     & 1      & 1,203 \\
& Asinh(Experience in the field) & 2.705 & 1.207 & 0     & 4.820  & 1,203 \\
\hline

$\phi = 5.5$
& Asinh(Production)              & 5.571 & 1.422 & 1.397 & 9.423  & 991 \\
& Asinh(Hit Non clinical)        & 5.016 & 2.383 & 0     & 9.256  & 991 \\
& Male                           & 0.752 & 0.432 & 0     & 1      & 991 \\
& Asinh(Experience in the field) & 2.656 & 1.237 & 0     & 4.820  & 991 \\
\hline

\end{tabular}
}
\end{table}

\section{Results} \label{sec:res}
We begin our empirical analysis by examining whether sharing a common cultural background predicts collaboration in both the scientific and technological networks. This analysis constitutes the formal first step of the empirical strategy outlined in Section \ref{sec:econ}. We then turn to the second step, which investigates the interaction between scientific productivity and inventive output.

\subsection{Relevance of instruments}
The results from the dyadic network formation models for both the scientific and technological domains reveal a set of consistent patterns, underscoring the parallel mechanisms shaping collaboration across research and invention.

In Tables \ref{tab:log1} and \ref{tab:log2}, the results of the logistic regression for the network formation model can be seen. To begin the analysis, we examine the exclusion restriction. Sharing a common cultural background, broadly defined, appears relevant, as evidenced by a statistically significant coefficient with the expected sign. This suggests that two agents with similar cultural background tend to have a higher collaboration intensity. This finding aligns with existing literature that highlights the importance of shared cultural background in driving collaboration formation.

Social and institutional proximity emerge as the dominant predictors of collaboration intensity. Proximity within prior collaborative structures, whether through coauthorship or coinventorship, significantly increases the likelihood of subsequent collaboration, particularly at distances of one or two. This finding echoes the idea that existing ties provide a channel for information about the reliability, productivity, and complementary expertise of potential partners, thereby reducing uncertainty and coordination costs in team formation.

Institutional proximity exhibits a similar and robust effect across domains. In medical and technological research alike, collaboration is facilitated by shared institutional environments which reduce logistical frictions and enable easier access to specialized infrastructures such as laboratories, hospitals, or proprietary equipment. Co-location not only lowers the cost of interaction but also fosters trust and informal communication, which often evolve into formal collaborations over time. These patterns support the view that institutions play a central role in structuring both scientific and inventive teamwork.

Finally, gender homophily is also present in both activities, although the estimated effect is weaker in the technological network. In our case, we are unable to determine through which specific channels gender homophily operates, whether through social preferences, network segregation, or differences in risk attitudes, leaving this as an interesting avenue for future research. 

Overall, these results reveal striking similarities between the determinants of collaboration in science and technology. Cultural, social, institutional, and gender proximities jointly shape the formation of both coauthorship and coinventorship links, underscoring the pervasive role of interpersonal and organizational structures in sustaining knowledge production across domains.

\begin{table}[!htbp] 
\centering 
\caption{\label{tab:log1} Estimates of the coefficients of the logistic regression in the network formation model for the technology network $\bm{G}_T$.} 
\begin{tabular}{@{\extracolsep{5pt}}lc} 
\\[-1.8ex]\hline 
\hline \\[-1.8ex] 
 & \multicolumn{1}{c}{\textit{Dependent variable: $(g_{ij})_{T} = 1$}} \\ 
\\[-1.8ex]  
\hline \\[-1.8ex] 
Same Ethnicity & 0.259$^{***}$ \\ 
  & (0.036) \\ 
Asinh (Diff experience)  & $-$0.014$^{***}$ \\ 
  & (0.003) \\ 
Past Coinventor & 1.670$^{***}$ \\ 
  & (0.069) \\ 
Past Common coinventor & 0.283$^{***}$ \\ 
  & (0.054) \\ 
Same gender & $-$0.065$^{*}$ \\ 
  & (0.037) \\ 
Same Institution & 1.316$^{***}$ \\ 
  & (0.049) \\ 
\hline \\[-1.8ex] 
Observations & 30,870 \\ 
Log Likelihood & $-$10,679.730 \\ 
Akaike Inf. Crit. & 21,373.450 \\ 
McFadden Pseudo $R^{2}$ & 0.11897\\
\hline 
\end{tabular} 
\begin{minipage}{\textwidth}
\footnotesize \textit{Note} The dependent variable is defined as the existence of collaboration between inventor $i$ and inventor $j$. The independent variables capture differences in characteristics between $i$ and $j$. A precise definition of the variables at the individual level can be found in Section \ref{sec:data}. Standard errors are reported in parentheses. An intercept is included. ${*}$, ${**}$, and ${***}$ indicate statistical significance at the 10\%, 5\%, and 1\% levels.
\end{minipage}
\end{table} 

\begin{table}[H] 
\centering 
\caption{\label{tab:log2} Estimates of the coefficients of the logistic regression in the network formation model for the science network $\bm{G}_S$.} 
\begin{tabular}{@{\extracolsep{5pt}}lcccc} 
\\[-1.8ex]\hline 
\hline \\[-1.8ex] 
& \multicolumn{4}{c}{\textit{Dependent variable: $(g_{ij})_{S} = 1$}} \\
& $\psi = 4$  & $\psi = 4.5$ & $\psi = 5$   & $\psi = 5.5$ \\
\hline \\[-1.8ex] 
Cosine Similarity & 2.138$^{***}$ & 2.438$^{***}$ & 1.905$^{***}$ & 2.277$^{***}$ \\ 
& (0.167) & (0.183) & (0.172) & (0.194) \\ 
Same Ethnicity & 0.357$^{***}$ & 0.315$^{***}$ & 0.306$^{***}$ & 0.280$^{***}$ \\ 
& (0.039) & (0.042) & (0.040) & (0.045) \\ 
Same Gender & 0.128$^{***}$ & 0.133$^{***}$ & 0.154$^{***}$ & 0.105$^{**}$ \\ 
& (0.039) & (0.042) & (0.040) & (0.045) \\ 
Same Institution & 1.957$^{***}$ & 1.816$^{***}$ & 1.915$^{***}$ & 1.775$^{***}$ \\ 
& (0.042) & (0.046) & (0.043) & (0.050) \\ 
Shared Past Coauthors & 0.883$^{***}$ & 0.937$^{***}$ & 0.897$^{***}$ & 0.905$^{***}$ \\ 
& (0.049) & (0.051) & (0.049) & (0.054) \\ 
\hline \\[-1.8ex] 
Observations & 49,571 & 37,773 & 44,440 & 31,647 \\ 
Log Likelihood & $-$10,666.890 & $-$9,044.039 & $-$10,281.630 & $-$8,002.431 \\ 
Akaike Inf. Crit. & 21,379.780 & 18,134.080 & 20,609.260 & 16,048.860 \\ 
McFadden Pseudo $R^{2}$ & 0.1460 & 0.1288 & 0.1335 &  0.1198 \\ 
\hline \\[-1.8ex]  
\end{tabular} 
\begin{minipage}{\textwidth}
\footnotesize \textit{Note} The dependent variable is defined as the existence of collaboration betwen author $i$ and author $j$. The independent variables capture differences in characteristics between $i$ and $j$. A precise definition of the variables at the individual level can be found in Section \ref{sec:data}. Standard errors are reported in parentheses. An intercept is included. ${*}$, ${**}$, and ${***}$ indicate statistical significance at the 10\%, 5\%, and 1\% levels.
\end{minipage}
\end{table} 

\subsection{Effects of the scientific efforts on inventive productivity}
Regarding the influence of inventor characteristics on research output, we find that the cumulative novelty generated by an inventor’s past work is a positive and significant predictor of current productivity. This result is consistent with prior studies \citep{singh2010lone, ductor2015does}, which show that past impact, measured as output weighted by its degree of novelty, strongly correlates with sustained inventive performance. In contrast, seniority, proxied by the number of decades since an inventor’s first patent application, exerts a negative effect on research output. This finding echoes the evidence in \cite{ductor2015does}, who documents a decline in productivity with career length, and aligns with the life-cycle patterns identified by \cite{levin1991research}, suggesting that scientific output tends to decrease as researchers progress through their careers.

Columns 1 to 8 of Table \ref{tab:res_tech} present the estimate of the simultaneous equation model using both 2SLS-1 and 2SLS-2 controlling for network endogeneity. As evidence by the Over Identification restriction test, our estimator allows to succefully correct the network endogeneity.

The coinventor parameter $\lambda_{T}$ being positive and significant supports the strategic complementarity argument according to which an inventor which is collaborating with more productive coinventors have an incentive to also increase their productivity.\\
Turning to the main parameter of interest, $\lambda_{ST}$, we find that it is positive and statistically significant across all specifications. This indicates that an increase in scientific production is associated with a contemporaneous rise in inventive productivity. In other words, greater research output generates incentives or complementarities that stimulate innovation activity. It is important to note that this effect captures simultaneous, rather than lagged, interactions between scientific and technological production. This finding complements previous studies that document positive spillovers of scientific activity on subsequent patenting outcomes by showing that the effect originates during the collaborative process itself, rather than being solely explained by the traditional "paper trail" argument linking publications to later inventions.

\begin{table}[H]
\caption{\label{tab:res_tech} Estimates of the coefficients and structural parameters in the interrelated simultaneous equation network model. Results are reported from the perspective of the science network $\bm{G}_S$.}
\centering
\resizebox{\textwidth}{!}{%
\begin{tabular}{|l|cc|cc|cc|cc|} 
\hline
\textbf{Variable} & \multicolumn{2}{c|}{\bm{$\psi = 4$}} & \multicolumn{2}{c|}{\bm{$\psi = 4.5$}} & \multicolumn{2}{c|}{\bm{$\psi = 5$}} & \multicolumn{2}{c|}{\bm{$\psi = 5.5$}} \\ 
& \textbf{2SLS} & \textbf{2SLS-EC} & \textbf{2SLS} & \textbf{2SLS-EC} & \textbf{2SLS} & \textbf{2SLS-EC} & \textbf{2SLS} & \textbf{2SLS-EC} \\
\hline 
$\lambda_{T}$&0.0331$^{***}$&0.0322$^{***}$&0.0331$^{***}$&0.0322$^{***}$&0.0331$^{***}$&0.0322$^{***}$&0.0331$^{***}$&0.0322$^{***}$ \\
&(0.0019)&(0.0044)&(0.0019)&(0.0044)&(0.0019)&(0.0044)&(0.0019)&(0.0044) \\ $\lambda_{ST}$&0.0509$^{**}$&0.0507$^{**}$&0.0509$^{**}$&0.0507$^{**}$&0.0509$^{**}$&0.0507$^{**}$&0.0509$^{**}$&0.0507$^{**}$ \\
&(0.0226)&(0.0227)&(0.0226)&(0.0227)&(0.0226)&(0.0227)&(0.0226)&(0.0227) \\
Citations &0.0826$^{***}$&0.0832$^{***}$&0.0826$^{***}$&0.0832$^{***}$&0.0826$^{***}$&0.0832$^{***}$&0.0826$^{***}$&0.0832$^{***}$ \\
&(0.0085)&(0.0092)&(0.0085)&(0.0092)&(0.0085)&(0.0092)&(0.0085)&(0.0092) \\ Gender&0.0315$^{}$&0.0311$^{}$&0.0315$^{}$&0.0311$^{}$&0.0315$^{}$&0.0311$^{}$&0.0315$^{}$&0.0311$^{}$ \\
&(0.0451)&(0.0452)&(0.0451)&(0.0452)&(0.0451)&(0.0452)&(0.0451)&(0.0452) \\ Experience &-0.0077$^{**}$&-0.008$^{**}$&-0.0077$^{**}$&-0.008$^{**}$&-0.0077$^{**}$&-0.008$^{**}$&-0.0077$^{**}$&-0.008$^{**}$ \\
&(0.0038)&(0.0041)&(0.0038)&(0.0041)&(0.0038)&(0.0041)&(0.0038)&(0.0041) \\ 
\hline

\hline 
\text{OIR test p-value} &0.1275&0.194&0.1275&0.194&0.1275&0.194&0.1275&0.194 \\ 
\text{Cragg-Donald Wald F-statistic} &103.2422&54.0867&103.2422&54.0867&103.2422&54.0867&103.2422&54.0867 \\ 
Communities fixed effects&YES&YES&YES&YES&YES&YES&YES&YES \\
\hline 
\end{tabular}
}
 \begin{minipage}{\textwidth}
    \footnotesize 
    \textbf{Notes:} Estimates are derived for several resolution parameter  $\psi \in \{4,4.5,5,5.5\}$ for the Leiden algorithm. For columns 1–8, the dependent variable is computed using the inverse hyperbolic sine (asinh) of the patent production weighted by its novelty defined by \cite{kelly2021measuring}. The scientific output needed to compute $\lambda_{ST}$ is computed using asinh of the total number of citations over the study period, weighted by journal impact factor. The following dummy variables is included: Male. Inventor's experience is captured by the number of years since the first application. Citations is computed using asinh of the sum of citations received before the beginning of the period under consideration. Heteroskedasticity-robust standard errors are reported in parentheses. Statistical significance is indicated by ${*}$, ${**}$, and ${***}$ for the 10\%, 5\%, 
    and 1\% levels, respectively.
    \end{minipage}
\end{table}

\subsection{Effects of the inventive efforts on scientific productivity}
Consistent with the patterns observed for inventors, past research impact, measured as cumulative publication output weighted by citations and the impact factor of the citing journals, is a strong and significant predictor of current productivity. This aligns with prior evidence in \cite{ductor2015does} that high-impact scientists tend to remain highly productive, reflecting the self-reinforcing nature of scientific influence. In contrast, seniority, proxied by the number of decades since a researcher’s first publication, exerts a negative effect on output, echoing the life-cycle patterns documented by \cite{levin1991research}.

Table \ref{tab:res2} presents the estimates of the simultaneous equation network model using both 2SLS and 2SLS-EC controlling for network endogeneity. As evidence by the Over Identification restriction test, $\bm{G}_{S}$ is most likely endogenous as evidence by the low OIR p-value. However, we are able to successfully correct the network endogeneity by using our estimator.

Finally, consistent with the results from the inventor network analysis, the coauthorship parameter $\lambda_{S}$ is positive and significant, indicating that collaboration with more productive peers generates strategic complementarities that enhance a scientist’s own output. Together, these findings mirror the complementarity patterns observed in the technological network.

The estimated coefficient $\lambda_{TS}$ on technological productivity is negative but statistically insignificant, indicating that current inventive activity does not appear to influence scientific productivity. This finding supports the notion that the complementarity between scientific and technological productivity is asymmetric: scientific output clearly influences contemporaneous inventive activity, whereas the reverse effect cannot be supported. In line with the linear model of the science–industry relationship, scientific research underpins technological development, with discoveries providing the foundation for subsequent innovation. Our results reinforce this framework, showing that while scientific production stimulates contemporaneous inventive activity ($\lambda_{ST}$), inventive activity does not have a measurable impact on ongoing scientific research.

\begin{table}[H]
\caption{\label{tab:res2} Estimates of the coefficients and structural parameters in the interrelated simultaneous equation network model. Results are reported from the perspective of the technology network $\bm{G}_T$.}
\resizebox{\textwidth}{!}{%
\centering
\begin{tabular}{|l|cc|cc|cc|cc|} 
\hline
\textbf{Variable} & \multicolumn{2}{c|}{$\bm{\psi} = 4$} & \multicolumn{2}{c|}{$\bm{\psi} = 4.5$} & \multicolumn{2}{c|}{$\bm{\psi} = 5$} & \multicolumn{2}{c|}{$\bm{\psi} = 5.5$} \\ 
& \textbf{2SLS} & \textbf{2SLS-EC} & \textbf{2SLS} & \textbf{2SLS-EC} & \textbf{2SLS}& \textbf{2SLS-EC} & \textbf{2SLS} & \textbf{2SLS-EC} \\
\hline 
$\lambda_{S}$&0.0079$^{***}$&0.0117$^{***}$&0.0076$^{***}$&0.0109$^{***}$&0.0071$^{***}$&0.0102$^{***}$&0.0079$^{***}$&0.0079$^{*}$ \\
&(0.0013)&(0.0035)&(0.0013)&(0.004)&(0.0012)&(0.0038)&(0.0015)&(0.0045) \\ $\lambda_{TS}$&-0.135$^{}$&-0.1286$^{}$&-0.1348$^{}$&-0.1292$^{}$&-0.118$^{}$&-0.111$^{}$&-0.1182$^{}$&-0.1223$^{}$ \\
&(0.1035)&(0.1034)&(0.1053)&(0.1056)&(0.106)&(0.1058)&(0.1064)&(0.1069) \\ Asinh(Past Production non clinical) &0.2212$^{***}$&0.2102$^{***}$&0.216$^{***}$&0.2085$^{***}$&0.2491$^{***}$&0.2407$^{***}$&0.2255$^{***}$&0.2255$^{***}$ \\
&(0.0276)&(0.0287)&(0.0292)&(0.0298)&(0.0279)&(0.0286)&(0.031)&(0.0317) \\ Gender &0.1231$^{}$&0.1157$^{}$&0.1195$^{}$&0.1159$^{}$&0.1467$^{*}$&0.1376$^{}$&0.1538$^{}$&0.154$^{}$ \\
&(0.0852)&(0.0855)&(0.0882)&(0.0885)&(0.0875)&(0.0882)&(0.096)&(0.0959) \\ Experience &-0.1582$^{***}$&-0.1549$^{***}$&-0.141$^{***}$&-0.1381$^{**}$&-0.2154$^{***}$&-0.2123$^{***}$&-0.1855$^{***}$&-0.1854$^{***}$ \\
&(0.0521)&(0.052)&(0.054)&(0.0538)&(0.0523)&(0.0521)&(0.0573)&(0.0572) \\ 
\hline

\hline 
\text{OIR test p-value} &0.0338&0.518&0.0581&0.6744&0.0339&0.9135&0.1015&0.6608 \\ 
\text{Cragg-Donald Wald F-statistic} &33.2578&26.9579&27.7835&22.155&33.2305&27.3607&25.8795&18.9682 \\ 
Communities fixed effects&YES&YES&YES&YES&YES&YES&YES&YES \\
\hline
\end{tabular}
}
 \begin{minipage}{\textwidth}
    \footnotesize 
    \textbf{Notes:} Estimates are derived for several resolution parameter $\psi \in \{4,4.5,5,5.5\}$ for the Leiden algorithm. For columns 1–8, the dependent variable is computed using asinh of the total number of citations over the study period, weighted by journal impact factor. The technological output needed to compute $\lambda_{TS}$  is computed using asinh of the patent production weighted by its novelty defined by \cite{kelly2021measuring}. The following dummy variables is included: Male. Author's experience is captured by the number of years since the first publication in the field. \textit{Hit} is computed using asinh of the maximum citations received for a non clinical research article. Heteroskedasticity-robust standard errors are reported in parentheses. Statistical significance is indicated by ${*}$, ${**}$, and ${***}$ for the 10\%, 5\%, 
    and 1\% levels, respectively.
    \end{minipage}
\end{table}


\subsection{Robustness check}
While all models reported in Table~\ref{tab:res_tech} and Table~\ref{tab:res2} control for each individual's past production, one may still be concerned that the positive impact of academia–industry collaboration on inventors' productivity is driven by the selection of more productive inventors into science. In other words, if inventors with higher patenting productivity were more likely to become authors, and if these same inventors were also inherently more productive, our estimates of the effect of dual involvement could be biased.\\
To address this concern, we estimate a series of logistic regressions in which the dependent variable is the likelihood of being an author, and the explanatory variables include each inventor's past patenting productivity, patenting experience, past scientific productivity, and gender. The results, reported in Table~\ref{tab:rob1}, show that participation in science is positively associated with past scientific productivity. Crucially, the estimates indicate that the likelihood of becoming an author is not correlated with past patenting productivity or with patenting experience. This finding suggests that more productive inventors are not self selecting into science, which alleviates concerns that our main results are driven by self-selection of high-performing inventors into science.
\begin{table}[H] \centering 
  \caption{Is being an author conditioned by being a good inventor?} 
  \label{tab:rob1} 
\begin{tabular}{@{\extracolsep{5pt}}lcccc} 
\\[-1.8ex]\hline 
\hline \\[-1.8ex] 
 & \multicolumn{4}{c}{\textit{Dependent variable:}} \\ 
\\[-1.8ex] & \multicolumn{4}{c}{Being an author} \\ 
\\[-1.8ex] & $\psi = 4$ & $\psi = 4.5$ & $\psi = 5$ & $\psi = 5.5$\\ 
\hline \\[-1.8ex] 
 Asinh(Past Production Kelly) & $-$0.223 & $-$0.223 & $-$0.223 & $-$0.223 \\ 
  & (0.207) & (0.207) & (0.207) & (0.207) \\ 
 Asinh(Experience Patenting) & 0.386 & 0.386 & 0.386 & 0.386 \\ 
  & (0.250) & (0.250) & (0.250) & (0.250) \\ 
 Male & 0.293 & 0.293 & 0.293 & 0.293 \\ 
  & (0.474) & (0.474) & (0.474) & (0.474) \\ 
 Asinh(Past Citations Science) & 0.481$^{***}$ & 0.481$^{***}$ & 0.481$^{***}$ & 0.481$^{***}$ \\ 
  & (0.072) & (0.072) & (0.072) & (0.072) \\ 
 Constant & $-$6.002$^{***}$ & $-$6.002$^{***}$ & $-$6.002$^{***}$ & $-$6.002$^{***}$ \\ 
  & (0.702) & (0.702) & (0.702) & (0.702) \\ 
\hline \\[-1.8ex] 
Observations & 947 & 947 & 947 & 947 \\ 
Log Likelihood & $-$114.162 & $-$114.162 & $-$114.162 & $-$114.162 \\ 
Akaike Inf. Crit. & 238.323 & 238.323 & 238.323 & 238.323 \\ 
\hline 
\hline \\[-1.8ex] 
\end{tabular}

\begin{minipage}{\textwidth}
\footnotesize \textit{Note} The dependent variable is defined as being an author for an inventor. The independent variables capture past patenting productivity measure by past patent production weighted by \cite{kelly2021measuring} novelty index, experience in patenting is measured in years starting from the first application, past scientific productivity is captured by the past scientific production weighted by citations, and a dummy variable for being a male is included. An intercept is included. ${*}$, ${**}$, and ${***}$ indicate statistical significance at the 10\%, 5\%, and 1\% levels.
\end{minipage}
\end{table} 
The results, reported in Table~\ref{tab:rob2}, show that participation in technology is not significantly associated with past scientific productivity. Crucially, the estimates indicate that the likelihood of becoming an inventor is not correlated with past scientific productivity or with scientific experience. This finding suggests that more productive authors are not self selecting into technology, which alleviates concerns that our main results are driven by self-selection of high-performing authors into technology.
\begin{table}[H] \centering 
  \caption{Is being an inventor conditionned by being a productive author?} 
  \label{tab:rob2} 
\begin{tabular}{@{\extracolsep{5pt}}lcccc} 
\\[-1.8ex]\hline 
\hline \\[-1.8ex] 
 & \multicolumn{4}{c}{\textit{Dependent variable:}} \\ 
\\[-1.8ex] & \multicolumn{4}{c}{Being inventor} \\ 
\\[-1.8ex] & $\psi = 4$ & $\psi = 4.5$ & $\psi = 5$ & $\psi = 5.5$\\ 
\hline \\[-1.8ex] 
 Asinh(Hit Non clinical) & $-$0.053 & $-$0.053 & $-$0.056 & $-$0.040 \\ 
  & (0.087) & (0.086) & (0.088) & (0.088) \\ 
 Experiences in the field & 0.151 & 0.163 & 0.171 & 0.175 \\ 
  & (0.197) & (0.190) & (0.197) & (0.197) \\ 
 Constant & $-$3.402$^{***}$ & $-$3.303$^{***}$ & $-$3.381$^{***}$ & $-$3.268$^{***}$ \\ 
  & (0.390) & (0.392) & (0.391) & (0.391) \\ 
\hline \\[-1.8ex] 
Observations & 1,236 & 1,111 & 1,203 & 991 \\ 
Log Likelihood & $-$165.959 & $-$161.886 & $-$164.873 & $-$157.567 \\ 
Akaike Inf. Crit. & 337.919 & 329.773 & 335.745 & 321.134 \\ 
\hline 
\hline 
\hline \\[-1.8ex] 
\end{tabular}
\begin{minipage}{\textwidth}
\footnotesize \textit{Note} The dependent variable is defined as being an inventor for an author. The independent variables capture past scientific productivity measure by the asinh of the research non clinical paper which receives the highest number of citations, experience in the field is measured in years starting from the first publication in the field of cancer. ${*}$, ${**}$, and ${***}$ indicate statistical significance at the 10\%, 5\%, and 1\% levels.
\end{minipage}
\end{table}

\section{Conclusion} \label{sec:con}
In this paper, we develop a theory of scientific peer effects to study knowledge production when agents are influenced not only by the behavior of their peers but also by the bundles of activities in which they are jointly engaged. The model predicts that an agent’s productivity increases with their Katz–Bonacich centrality, a standard network centrality measure capturing both direct and indirect connections. We test these predictions using a rich and unique data set of researchers and inventors active in cancer research, a domain where scientific and technological activities are deeply intertwined.

Controlling for observable characteristics and unobserved network-specific heterogeneity, our empirical analysis indicates that an agent’s position in both collaboration networks, captured by a variant of Katz–Bonacich centrality, plays a central role in determining productivity. A central insight of our study is that, when examining the direct interaction between scientific and technological activities, the evidence supports the linear model of the science–industry relationship: scientific production has a contemporaneous positive effect on inventive output, whereas the reverse effect is negligible. This asymmetric pattern suggests that scientific knowledge remains the primary input into technological development, even within highly collaborative, industry-linked research environments. Understanding how immediate peers and broader network structures shape productivity thus has important implications for policies aimed at fostering knowledge creation and sustaining long-term growth.

Several avenues for future research emerge from our framework. First, it would be valuable to explore heterogeneity in peer effects, allowing them to vary with individual characteristics or field-specific attributes. Second, our model relies on a linear formulation of peer influence; relaxing this assumption by allowing for nonlinear or threshold-based spillovers may generate richer implications. Finally, although we highlight the policy relevance of our findings, a systematic analysis of the model’s policy implications has not been conducted and represents a natural next step for future research.

\backmatter

\bmhead{Supplementary information}
All R-code for the reproducibility of the empirical results as well as the implemented estimation strategy along with the Monte Carlo experiments is available upon request.

\section*{Declarations}

\bmhead{Funding}
Financial support by the German Research Foundation (DFG) [RTG 2865/1 – 492988838] is gratefully acknowledged.

\bmhead{Conflict of interests}
The authors declare that they have no conflict of interest.

\newpage
\appendix

\section{Social Planner's Optimization Problem} \label{app:plan}
The social planner’s objective is to maximize the total sum of utilities from both science and technology activities across all agents in the network. The utility of agent $i$ is given by
\begin{align*}
   u_i(\mathbf{y}_i, \mathbf{y}_{-i}) &= y_i^{(T)} \alpha_i^{(T)} + y_i^{(S)} \alpha_i^{(S)} - \frac{1}{2} (y_i^{(T)})^2 - \frac{1}{2} (y_i^{(S)})^2 - \beta y_i^{(S)} y_i^{(T)} \\
   &+ \lambda_T \sum_{j \in N_T} g_{ij}^{(T)} y_i^{(T)} y_j^{(T)} + \lambda_S \sum_{j \in N_S} g_{ij}^{(S)} y_i^{(S)} y_j^{(S)}. 
\end{align*}
In this case, $y_i^{(T)}$ and $y_i^{(S)}$ represent the technology and science production levels of agent $i$, respectively, and $\alpha_i^{(T)}$ and $\alpha_i^{(S)}$ represent their intrinsic productivity in the two activities. The parameter $\beta \in \mathbb{R}$ captures the interdependence between the two activities, while $\lambda_T$ and $\lambda_S$ represent the network effects on technology and science, respectively. The term $g_{ij}^{(T)}$ represents the network adjacency matrix for technology collaborations, and similarly, $g_{ij}^{(S)}$ represents the network adjacency matrix for science collaborations. The planner maximizes the sum of utilities given by
\begin{equation*}
    \sum_{i \in N} u_i(\mathbf{y}_i, \mathbf{y}_{-i}).
\end{equation*}
To find the optimal production levels, we first derive the first-order conditions (FOC) by differentiating the total utility function with respect to $y_i^{(T)}$ and $y_i^{(S)}$. The derivative of $u_i$ with respect to $y_i^{(T)}$ yields
\begin{equation*}
    \frac{\partial u_i}{\partial y_i^{(T)}} = \alpha_i^{(T)} - y_i^{(T)} - \beta y_i^{(S)} + \lambda_T \sum_{j \in N_T} g_{ij}^{(T)} y_j^{(T)} = 0.
\end{equation*}
Similarly, the derivative of $u_i$ with respect to $y_i^{(S)}$ yields
\begin{equation*}
  \frac{\partial u_i}{\partial y_i^{(S)}} = \alpha_i^{(S)} - y_i^{(S)} - \beta y_i^{(T)} + \lambda_S \sum_{j \in N_S} g_{ij}^{(S)} y_j^{(S)} = 0.  
\end{equation*}
The planner’s goal is to maximize the total utility, which involves solving for the optimal production levels $y_i^{(T)}$ and $y_i^{(S)}$ that satisfy the above FOC across all agents. The planner's objective function, summing over all agents, is then given as
\begin{align*}
    \sum_{i \in N}  y_i^{(T)} \alpha_i^{(T)} &+ y_i^{(S)} \alpha_i^{(S)} - \frac{1}{2} (y_i^{(T)})^2 - \frac{1}{2} (y_i^{(S)})^2 - \beta y_i^{(S)} y_i^{(T)} \\ 
    &+ \lambda_T \sum_{j \in N_T} g_{ij}^{(T)} y_i^{(T)} y_j^{(T)} + \lambda_S \sum_{j \in N_S} g_{ij}^{(S)} y_i^{(S)} y_j^{(S)} .
\end{align*}
We can express the system of FOC in matrix form. Let $\bm{y}^{(T)}$ and $\bm{y}^{(S)}$ be the vectors of production levels for technology and science, respectively. The planner’s system of equations becomes
\begin{equation*}
    \begin{pmatrix}
    \bm{I} - 2 \lambda_T \bm{G}_T & \beta \bm{G}_S \\
    \beta \bm{G}_T & \bm{I} - 2 \lambda_S \bm{G}_S
    \end{pmatrix}
    \begin{pmatrix}
    \bm{y}^{(T)} \\
    \bm{y}^{(S)}
    \end{pmatrix}
    =
    \begin{pmatrix}
    \bm{\alpha}^{(T)} \\
    \bm{\alpha}^{(S)}
    \end{pmatrix}   
\end{equation*}
To solve for the optimal production levels we solve the system of linear equations
\begin{equation*}
    \bm{y}^* = 
\begin{pmatrix}
\bm{I} - 2\lambda_T \bm{G}_T & -\beta \bm{G} _{ST} \\
-\beta \bm{G}_{TS} & \bm{I} - 2\lambda_S \bm{G}_S
\end{pmatrix}^{-1}
\begin{pmatrix}
\bm{\alpha}^{(T)} \\
\bm{\alpha}^{(S)}
\end{pmatrix}.
\end{equation*}
This gives the optimal production levels $\bm{y}^{(T)*}$ and $\bm{y}^{(S)*}$ which maximize the total utility of all agents in the system, considering both network effects and the interdependence between science and technology activities.

\section{Two-stage least squares calculation process} \label{app:2sls}
\subsection{Assumptions}
Let $\bm{y} = \begin{pmatrix}
    \bm{y}^{(T)} \\
    \bm{y}^{(S)}
\end{pmatrix}$, 
$\bm{G} = \begin{bmatrix}
    \lambda_T \bm{G}_{T} & \lambda_{TS}  \bm{G}_{TS} \\
    \lambda_{ST}  \bm{G}_{ST} & \lambda_S \bm{G}_{S}
\end{bmatrix}$,
$\bm{\mu} = \begin{pmatrix}
    \bm{\mu}_{T} \\
    \bm{\mu}_{S}
\end{pmatrix}$,
$\bm{L} = \begin{bmatrix}
    \bm{L}_T & 0 \\
    0 & \bm{L}_S
\end{bmatrix}$,
$\bm{X} = \begin{bmatrix}
    \bm{X}_T & 0 \\
    0 & \bm{X}_S
\end{bmatrix}$, $\bm{\gamma} = \begin{pmatrix}
    \bm{\gamma}_T \\
    \bm{\gamma}_S
\end{pmatrix}$ and $\bm{\epsilon} = \begin{pmatrix}
    \bm{\epsilon}_T \\
    \bm{\epsilon}_S
\end{pmatrix}$, then the reduced form of the econometric network model is given by 
\begin{align*}
    \bm{y} &= \bm{G} \bm{y} + \bm{L}\bm{\mu} +\bm{X}\bm{\gamma} + \bm{\epsilon} \\
    \bm{y} &= (\bm{I} - \bm{G})^{-1} \bm{L}\bm{\mu} + (\bm{I} - \bm{G})^{-1} \bm{X}\bm{\gamma}  + (\bm{I} - \bm{G})^{-1} \bm{\epsilon} \\
    \bm{y} &= \bm{S}^{-1}(\bm{L}\bm{\mu} + \bm{X}\bm{\gamma} + \bm{\epsilon})
\end{align*}
where $\bm{S} = (\bm{I} - \bm{G})$. Following regularity conditions are imposed for the derivation of the asymptotic properties of the 2SLS estimator.

\begin{assumption} \label{ass:error}
    The vector of idiosyncratic random innovations is given by $\bm{\epsilon} = (\bm{\epsilon}_T', \bm{\epsilon}_S')' = \left(\left(\bm{\epsilon}^{(T)\prime}_{1}, \dots, \bm{\epsilon}^{(T)\prime}_{\bar{c}_T}\right)' , \left(\bm{\epsilon}^{(S)\prime}_{1}, \dots, \bm{\epsilon}^{(S)\prime}_{\bar{c}_S}\right)'\right)$ such that the vectors $\left\{\bm{\epsilon}^{(T)}_{c_T}\right\}_{c_T=1}^{\bar{c}_T}$ and $\left\{\bm{\epsilon}^{(S)}_{c_S}\right\}_{c_S=1}^{\bar{c}_S}$ are independent across $c_T$ and $c_S$. All disturbances have zero mean $\mathbb{E}\left[\bm{\epsilon}^{(T)}_{c_T}\right] = \bm{0}$, $\mathbb{E}\left[\bm{\epsilon}^{(S)}_{c_S}\right] = \bm{0}$. Heteroskedasticity is allowed across observational units $\mathbb{E}\left[\bm{\epsilon}^{(T)}_{c_T}\bm{\epsilon}^{(T)\prime}_{c_T}\right] = \bm{\Sigma}_{T,c_T}$, $\mathbb{E}\left[\bm{\epsilon}^{(S)}_{c_S}\bm{\epsilon}^{(S)\prime}_{c_S}\right] = \bm{\Sigma}_{S,c_S}$
    where $\bm{\Sigma}_{T,c_T}$ and $\bm{\Sigma}_{T,c_S}$ are positive definite and may vary across $c_T$ and $c_S$. There is no cross-equation correlation $\text{Cov}\left(\bm{\epsilon}^{(T)}_{c_T}, \bm{\epsilon}^{(S)}_{c_S}\right) = \bm{0}
    \quad \forall c_T,c_S$. Consequently, the full covariance matrix of 
    $\bm{\epsilon}$ is block-diagonal $\text{Var}(\bm{\epsilon}) =
      \begin{pmatrix}
        \bm{\Sigma}_T & \mathbf{0}\\
        \mathbf{0} & \bm{\Sigma}_S
      \end{pmatrix}
      \equiv \bm{\Sigma}$
    where $\bm{\Sigma}_T = \text{diag}(\bm{\Sigma}_{T,1},\ldots,\bm{\Sigma}_{T,\bar{c}_T})$ and $\bm{\Sigma}_S = \text{diag}(\bm{\Sigma}_{S,1},\ldots,\bm{\Sigma}_{S,\bar{c}_S})$. The disturbances admit a square-root representation $\bm{\epsilon} = \bm{\Sigma}^{1/2} \bm{v},$ where $\bm{\Sigma}^{1/2} = \text{diag}(\bm{\Sigma}_T^{1/2}, \bm{\Sigma}_S^{1/2})$
    and the components of $\bm{v}$ are independent and identically distributed with mean zero, unit variance, and finite fourth moments. The eigenvalues of $\bm{\Sigma}_T$ and $\bm{\Sigma}_S$ are uniformly bounded by a finite constant.
\end{assumption}

\begin{assumption} \label{ass:col}
    The matrix of exogenous variables $\bm{X}$ has full column rank. Specifically, the limit $\lim_{n \to \infty} \bm{X}^{\prime} \bm{X}$ exists, is non-singular, and the elements of $\bm{X}$ are uniformly bounded in both row and column sums in absolute value.
\end{assumption}

\begin{assumption} \label{ass:mat}
    The sequences of matrices $\bm{\{G\}}$ and $\bm{\{S^{-1}\}}$ are uniformly bounded in absolute value and $\bm{S}$ is additionally non-singular.
\end{assumption}

\begin{assumption} \label{ass:FF}
    Let $\bm{F}_T = \mathbb{E}(\bm{J}_T \bm{Z}_T)$ and $\bm{F}_S = \mathbb{E}(\bm{J}_S \bm{Z}_S)$ such that $\bm{F} = \text{diag}\left(\bm{F}_T, \bm{F}_S\right)$. Then, $\bm{Q}_{TS} = \lim_{n \to \infty} \bm{F}_T'\bm{F}_S$, $\bm{Q}_{ST} = \lim_{n \to \infty} \bm{F}_S'\bm{F}_T$ and $\bm{Q} =  \lim_{n \to \infty} \bm{F}(\bm{\Sigma}^{-1} \otimes \bm{I}) \bm{F}$ are finite non-singular matrices
\end{assumption}

\begin{assumption} \label{ass:instruments}
    There are matrices $\bm{\Pi}_T$ and $\bm{\Pi}_S$ such that $|| \bm{F_T} - \bm{H}_{T,K} \bm{\Pi}_T||_{\infty} \to 0$ and $|| \bm{F_S} - \bm{H}_{S,K} \bm{\Pi}_S||_{\infty} \to 0$, as $K \to \infty$. 
\end{assumption}

Assumptions \ref{ass:error} to \ref{ass:mat} are based on \cite{kelejian2004} and ensure heteroskedasticity in idiosyncratic random innovations without cross-equation correlation, non-stochasticity of exogenous variables as well as the usual full rank condition and  the uniformly boundedness of $\bm{\{G\}}$ and $\bm{\{S^{-1}\}}$. Additionally, structural parameters obey the restrictions $|\lambda_T| + |\lambda_{TS}| < 1$ and $|\lambda_S| + |\lambda_{ST}| < 1$. The identification of the network is based on \cite{liu2010} and ensure that concentration parameter and sample size grow at the same rate which can be seen in Assumption \ref{ass:FF}. Since the best IV matrices are infeasible, Assumption \ref{ass:instruments} ensures that $\bm{F_T}$ and $\bm{F_S}$ can be approximated by a linear combination of feasible IV matrices as the number of IV increases with sample size.

\subsection{Instrumental variables}
Since $\mathbb{E}\left(\bm{G}_T \bm{y}^{(T)} \bm{\epsilon}_T'\right) \neq 0$, $\mathbb{E}\left(\bm{G}_{TS} \bm{y}^{(S)} \bm{\epsilon}_T'\right) \neq 0$, $\mathbb{E}\left(\bm{G}_S \bm{y}^{(S)} \bm{\epsilon}_S'\right) \neq 0$ and $\mathbb{E}\left(\bm{G}_{ST} \bm{y}^{(T)} \bm{\epsilon}_S'\right) \neq 0$, IV are necessary. However, the best IV matrices are infeasible due to the dependence on unknown parameters. Thus, the best IV matrices have to be approximated based on a linear combination of feasible IV matrices. Particularly, let $\bm{S}_T = (\bm{I}_{n_T} - \lambda_T \bm{G}_T)$ and $\bm{S}_S = (\bm{I}_{n_S} - \lambda_S \bm{G}_S)$ such that applying the partitioned inverse formula yields
\begin{equation*}
    S^{-1} = \begin{bmatrix}
    (\bm{S}_T^*)^{-1} & \lambda_{TS} \bm{S}_T^{-1} \bm{G}_{TS} (\bm{S}_S^*)^{-1} \\
     \lambda_{ST} \bm{S}_S^{-1} \bm{G}_{ST} (\bm{S}_T^*)^{-1} & (\bm{S}_S^*)^{-1}
\end{bmatrix}
\end{equation*}
where $\bm{S}_T^* = \bm{S}_T -  \lambda_{TS} \lambda_{ST} \bm{G}_{TS} \bm{S}_S^{-1} \bm{G}_{ST}$ and $\bm{S}_S^* =  \bm{S}_S -  \lambda_{ST} \lambda_{TS} \bm{G}_{ST} \bm{S}_T^{-1} \bm{G}_{TS}$. The equation of $\bm{y}^{(T)}$ contains the expressions $(\bm{S}_T^*)^{-1} \bm{X}_T$ and $\bm{S}_T^{-1} \bm{G}_{TS} (\bm{S}_S^*)^{-1} \bm{X}_S$. Utilizing the finite Neumann series approximation yields
\begin{align*}
    (\bm{S}_T^*)^{-1} \bm{X}_T &= (\bm{S}_T -  \lambda_{TS} \lambda_{ST} \bm{G}_{TS} \bm{S}_S^{-1} \bm{G}_{ST})^{-1} \bm{X}_T \\
    &= (\bm{I}_T -  \lambda_{TS} \lambda_{ST} \bm{S}_T^{-1} \bm{G}_{TS} \bm{S}_S^{-1} \bm{G}_{ST})^{-1} \bm{S}_T^{-1} \bm{X}_T \\
    &= (\bm{I}_T +  \lambda_{TS} \lambda_{ST} \bm{S}_T^{-1} \bm{G}_{TS} \bm{S}_S^{-1} \bm{G}_{ST} \\
    &+ \lambda_{TS}^2 \lambda_{ST}^2 (\bm{S}_T^{-1})^2 \bm{G}_{TS}^2 (\bm{S}_S^{-1})^2 \bm{G}_{ST}^2 + \cdots) \bm{S}_T^{-1} \bm{X}_T \\
    &= \bm{S}_T^{-1} \bm{X}_T  +  \lambda_{TS} \lambda_{ST} \bm{S}_T^{-1} \bm{G}_{TS} \bm{S}_S^{-1} \bm{G}_{ST} \bm{S}_T^{-1} \bm{X}_T  \\
    &+ \lambda_{TS}^2 \lambda_{ST}^2 (\bm{S}_T^{-1})^2 \bm{G}_{TS}^2 (\bm{S}_S^{-1})^2 \bm{G}_{ST}^2 \bm{S}_T^{-1} \bm{X}_T + \cdots \\
    &= (\bm{I}_{n_T} - \lambda_T \bm{G}_T)^{-1} \bm{X}_T + \lambda_{TS} \lambda_{ST} (\bm{I}_{n_T} - \lambda_T \bm{G}_T)^{-1} \\
    &\bm{G}_{TS} (\bm{I}_{n_S} - \lambda_S \bm{G}_S)^{-1} \bm{G}_{ST} (\bm{I}_{n_T} - \lambda_T \bm{G}_T)^{-1} \bm{X}_T + \cdots \\
    &= (\bm{I}_{n_T} + \lambda_T \bm{G}_T + \lambda_T^2 \bm{G}_T^2 + \cdots)^{-1} \bm{X}_T \\ 
    &+ \lambda_{TS} \lambda_{ST} (\bm{I}_{n_T} + \lambda_T \bm{G}_T + \lambda_T^2 \bm{G}_T^2 + \cdots)^{-1} \bm{G}_{TS} \\
    &(\bm{I}_{n_S} + \lambda_S \bm{G}_S + \lambda_S^2 \bm{G}_S^2 + \cdots)^{-1} \\ &\bm{G}_{ST} (\bm{I}_{n_T} + \lambda_T \bm{G}_T + \lambda_T^2 \bm{G}_T^2 + \cdots)^{-1} \bm{X}_T + \cdots 
\end{align*}
A similar expression can be derived for $\bm{S}_T^{-1} \bm{G}_{TS} (\bm{S}_S^*)^{-1} \bm{X}_S$. The matrix of feasible IV can thus be set to 
\begin{align*}
\bm{\tilde{H}}_{T,K} = \bm{J}_T
\big[&
\bm{X}_T,\,
\bm{G}_T\bm{X}_T,\,
\bm{G}_{TS}\bm{G}_{ST}\bm{X}_T,\,
\bm{G}_{TS}\bm{G}_{ST}\bm{G}_T\bm{X}_T,\,
\bm{G}_{TS}\bm{G}_S\bm{G}_{ST}\bm{X}_T,\, \\
&\bm{G}_T\bm{G}_{TS}\bm{G}_{ST}\bm{X}_T,\,
\bm{G}_{TS}\bm{G}_S\bm{G}_{ST}\bm{G}_T\bm{X}_T,\,
\bm{G}_T\bm{G}_{TS}\bm{G}_{ST}\bm{G}_T\bm{X}_T,\,\\
&\bm{G}_T\bm{G}_{TS}\bm{G}_S\bm{G}_{ST}\bm{X}_T,\, \bm{G}_T\bm{G}_{TS}\bm{G}_S\bm{G}_{ST}\bm{G}_T\bm{X}_T,\,
\bm{G}_{TS}\bm{X}_S,\,
\bm{G}_{TS}\bm{G}_S\bm{X}_S,\,\\
&\bm{G}_T\bm{G}_{TS}\bm{X}_S,\,
\bm{G}_T\bm{G}_{TS}\bm{G}_S\bm{X}_S,\,
\bm{G}_{TS}\bm{G}_{ST}\bm{G}_{TS}\bm{X}_S,\,
\bm{G}_{TS}\bm{G}_{ST}\bm{G}_{TS}\bm{G}_S\bm{X}_S,\,\\
&\bm{G}_{TS}\bm{G}_{ST}\bm{G}_T\bm{G}_{TS}\bm{X}_S,\,
\bm{G}_{TS}\bm{G}_S\bm{G}_{ST}\bm{G}_{TS}\bm{X}_S,\,
\bm{G}_T\bm{G}_{TS}\bm{G}_{ST}\bm{G}_{TS}\bm{X}_S,\,\\
&\bm{G}_{TS}\bm{G}_{ST}\bm{G}_T\bm{G}_{TS}\bm{G}_S\bm{X}_S,\,
\bm{G}_{TS}\bm{G}_S\bm{G}_{ST}\bm{G}_{TS}\bm{G}_S\bm{X}_S,\,\\
&\bm{G}_{TS}\bm{G}_S\bm{G}_{ST}\bm{G}_T\bm{G}_{TS}\bm{X}_S,\,
\bm{G}_T\bm{G}_{TS}\bm{G}_{ST}\bm{G}_{TS}\bm{G}_S\bm{X}_S,\,\\
&\bm{G}_T\bm{G}_{TS}\bm{G}_{ST}\bm{G}_T\bm{G}_{TS}\bm{X}_S,\,
\bm{G}_T\bm{G}_{TS}\bm{G}_S\bm{G}_{ST}\bm{G}_{TS}\bm{X}_S,\,\\
&\bm{G}_{TS}\bm{G}_S\bm{G}_{ST}\bm{G}_T\bm{G}_{TS}\bm{G}_S\bm{X}_S,\,
\bm{G}_T\bm{G}_{TS}\bm{G}_{ST}\bm{G}_T\bm{G}_{TS}\bm{G}_S\bm{X}_S,\\
&\bm{G}_T\bm{G}_{TS}\bm{G}_S\bm{G}_{ST}\bm{G}_{TS}\bm{G}_S\bm{X}_S,\,
\bm{G}_T\bm{G}_{TS}\bm{G}_S\bm{G}_{ST}\bm{G}_T\bm{G}_{TS}\bm{X}_S,\,\\
&\bm{G}_T\bm{G}_{TS}\bm{G}_S\bm{G}_{ST}\bm{G}_T\bm{G}_{TS}\bm{G}_S\bm{X}_S,\,
\cdots
\big]
\end{align*}
The derivation of $\bm{\tilde{H}}_{S,K}$ follows a similar manner. In practice, the number of IV can be reduced to avoid the weak instrument problem. For $\hat{\tilde{\bm{H}}}_{T,K}$ and $\hat{\tilde{\bm{H}}}_{S,K}$, the corresponding social adjacency matrix are replaced by the predicted counterparts $\bm{\hat{G}}_T$ and $\bm{\hat{G}}_S$ \citep{liu2014, baltagi2015, guo2019}. 

\subsection{Asymptotic properties}
The consistency of the 2SLS estimators can be quantified by observing that there is no dependence on the consistency of the estimators of the logistic regression $\hat{\bm{\tau}}^{(T)} = \left(\hat{\tau}_{0}^{(T)},\hat{\tau}_{1}^{(T)}\right)$, $\hat{\bm{\tau}}^{(S)} = \left(\hat{\tau}_{0}^{(S)},\hat{\tau}_{1}^{(S)}\right)$ as long as $\sqrt{n_T} \left(\hat{\bm{\tau}}^{(T)} - \bm{\tau}_{*}^{(T)}\right) = O_p(1)$ and $\sqrt{n_S} \left(\hat{\bm{\tau}}^{(S)} - \bm{\tau}_{*}^{(S)}\right) = O_p(1)$, that is, estimators converge to well defined limits $\bm{\tau}_{*}^{(T)}$ and $\bm{\tau}_{*}^{(S)}$. Let $\bm{G}^*_T$ and $\bm{G}^*_S$ be defined in a similar manner as $\bm{\hat{G}}_T$ and $\bm{\hat{G}}_S$. Then it follows under certain regularity conditions that $\hat{\tilde{\bm{H}}}_T^{'} \tilde{\bm{\epsilon}}_{T}(\bm{\delta}_T) = 0$ and $ \hat{\tilde{\bm{H}}}_S^{'} \tilde{\bm{\epsilon}}_{S} (\bm{\delta}_S) = 0$ are asymptotically identical to infeasible linear moment conditions $\tilde{\bm{H}}_T^{*'} \tilde{\bm{\epsilon}}_{T}(\bm{\delta}_T) = 0$ and $\tilde{\bm{H}}_S^{*'} \tilde{\bm{\epsilon}}_{S}(\bm{\delta}_S) = 0$ such that the 2SLS estimators are $\sqrt{n}$ consistent and asymptotically normal. The asymptotic distribution is then given by 
\begin{align*}
\sqrt{n_T} \left(\hat{\bm{\delta}}_T - \bm{\delta}_T\right)
&\xrightarrow{d} 
\mathcal{N}\left(0,
  \left(
    \begin{aligned}
      &\text{plim}_{n_T \to \infty} \frac{1}{n_T}
      \left(\bm{X}_T' \bm{P}_{\tilde{\bm{H}}_T^{*'}} \bm{X}_T\right)^{-1}
      \bm{X}_T' \bm{P}_{\tilde{\bm{H}}_T^{*'}} \bm{\Sigma}_T \\
      &\bm{P}_{\tilde{\bm{H}}_T^{*'}} \bm{X}_T
      \left(\bm{X}_T' \bm{P}_{\tilde{\bm{H}}_T^{*'}} \bm{X}_T\right)^{-1}
    \end{aligned}
  \right)^{-1}
\right)
\\
\sqrt{n_S} \left(\hat{\bm{\delta}}_S - \bm{\delta}_S\right)
&\xrightarrow{d} 
\mathcal{N}\left(0,
  \left(
    \begin{aligned}
      &\text{plim}_{n_S \to \infty} \frac{1}{n_S}
      \left(\bm{X}_S' \bm{P}_{\tilde{\bm{H}}_S^{*'}} \bm{X}_S\right)^{-1}
      \bm{X}_S' \bm{P}_{\tilde{\bm{H}}_S^{*'}} \bm{\Sigma}_S \\
      &\bm{P}_{\tilde{\bm{H}}_S^{*'}} \bm{X}_S
      \left(\bm{X}_S' \bm{P}_{\tilde{\bm{H}}_S^{*'}} \bm{X}_S\right)^{-1}
    \end{aligned}
  \right)^{-1}
\right)
\end{align*}

where $\bm{\Sigma}_T$ and $\bm{\Sigma}_S$ can be estimated via 
\begin{align*}
\hat{\bm{\Sigma}}_T
&=
\tilde{\bm{H}}_T^{*'} 
\operatorname{diag}\!\left(
\hat{\epsilon}_{T,1}^2, \hat{\epsilon}_{T,2}^2, \ldots, \hat{\epsilon}_{T,n_T}^2
\right)
\tilde{\bm{H}}_T^{*}, \\
\hat{\bm{\Sigma}}_S
&=
\tilde{\bm{H}}_S^{*'} 
\operatorname{diag}\!\left(
\hat{\epsilon}_{S,1}^2, \hat{\epsilon}_{S,2}^2, \ldots, \hat{\epsilon}_{S,n_S}^2
\right)
\tilde{\bm{H}}_S^{*},
\end{align*}
while $\hat{\bm{\epsilon}}_T = \tilde{\bm{y}}^{(T)} - \tilde{\bm{X}}_T \hat{\bm{\delta}}_T$ and $\hat{\bm{\epsilon}}_S = \tilde{\bm{y}}^{(S)} - \tilde{\bm{X}}_S \hat{\bm{\delta}}_S$ \citep{kelejian1998, lee2007, lee2021}.

\section{Monte Carlo experiments} \label{app:sim}
\subsection{Study Design}
To evaluate the performance of the proposed 2SLS and 2SLS-EC estimators, Monte Carlo experiments are conducted. Specifically, the focus is on the evaluation of the performance regarding the estimation of the 2SLS and 2SLS-EC estimators under weak and strong network effects for different strengths of endogeneity and fixed population and hence network sizes. Particularly, The true data generating process is given by
\begin{align*}
    \bm{y}^{(T)}  &= \lambda_T \bm{G}_T \bm{y}^{(T)} + \lambda_{TS} \bm{G}_{TS} \bm{y}^{(S)}  +  \bm{\mu}_{T} \bm{L}_T + \bm{\gamma}_T \bm{X}_{T} + \bm{\epsilon}_T  \\
    \bm{y}^{(S)} &= \lambda_S \bm{G}_{S} \bm{y}^{(S)} + \lambda_{ST} \bm{G}_{ST} \bm{y}^{(T)} + \bm{\mu}_{S} \bm{L}_S + \bm{\gamma}_S \bm{X}_{S} + \bm{\epsilon}_S.
\end{align*}
The number of communities is set to $\bar{c}_T = \bar{c}_S = 10$, each of equal size $n_{T,c_T} = n_{S,c_S} = 30$. The network fixed effects are drawn from the normal distribution $\bm{\mu}_T, \bm{\mu}_S  \sim 
\begin{bmatrix}
N(0,\bm{I}_{n_{T,c_T}}) \\
N(0,\bm{I}_{n_{S,c_S}})
\end{bmatrix}$ and unique characteristics of agents from uniform and normal distributions $\bm{X}_T, \bm{X}_S  \sim 
\begin{bmatrix}
\mathcal{U}(3,7) \\
N(0,0.5)
\end{bmatrix}$. The coefficients are set to $\bm{\gamma}_T = \bm{\gamma}_S= (1,0.5)^\prime$
Structural parameters of the networks are varied which capture effects of weak and strong network externalities as $\lambda_{T} \in \{0.3, 0.6\}$ and $\lambda_{S} \in \{0.2, 0.5\}$ while $\lambda_{TS} = \lambda_{ST} = 0.1$.

Furthermore, $\bm{G}_{T,c_T}$ and $\bm{G}_{S,c_T}$ are assumed to be endogenous and the networks are generated according to the probabilities of link formation via  
\begin{align*}
\mathbb{P}\left(\left(g_{ij, c_T}\right)_T = 1\right) &= \frac{\exp\left(\tau_{0}^{(T)} + \tau_{1}^{(T)} w_{ij,c_T}^{(T)} + u_{i,c_T}^{(T)} + u_{j,c_T}^{(T)}\right)}{1 + \exp\left(\tau_{0}^{(T)} + \tau_{1}^{(T)} w_{ij,c_T}^{(T)} + u_{i,c_T}^{(T)} + u_{j,c_T}^{(T)}\right)} \\
\mathbb{P}\left(\left(g_{ij, c_S}\right)_S  = 1\right) &= \frac{\exp\left(\tau_{0}^{(S)} + \tau_{1}^{(S)} w_{ij,c_S}^{(S)} + u_{i,c_S}^{(S)} + u_{j,c_S}^{(S)}\right)}{1 + \exp\left(\tau_{0}^{(S)} + \tau_{1}^{(S)} w_{ij,c_S}^{(S)} + u_{i,c_S}^{(S)} + u_{j,c_S}^{(S)}\right)}
\end{align*}
where observable similarity between agents is generated based on the second column of the unique characteristics of agents as $w_{ij,c_T}^{(T)} = 2 - (x_{i,c_T}^{(T)} - x_{j,c_T}^{(T)})^2$ and $w_{ij,c_S}^{(S)} = 2 - (x_{i,c_S}^{(S)} - x_{j,c_S}^{(S)})^2$. The corresponding coefficients are set to $\bm{\tau}^{(T)} = \bm{\tau}^{(S)} = (1,0.5)^\prime$. Unobserved heterogeneity is drawn based on the bivariate normal distribution as
\begin{equation*}
\begin{pmatrix} \nu_{i,c_T}^{(T)} \\ u_{i,c_T}^{(T)}\end{pmatrix}, \begin{pmatrix}  \nu_{i,c_S}^{(S)}\\ u_{i,c_S}^{(S)}\end{pmatrix} \sim \mathcal{N} \left( \begin{pmatrix} 0\\ 0\end{pmatrix}, \begin{pmatrix} 1 & \rho  \\ \rho & 1 \end{pmatrix} \right).
\end{equation*}
Endogeneity is then explicitly generated as $\epsilon^{(T)}_{i,c_T} = \nu_{i,c_T}^{(T)} \kappa_{i,c_T}^{(T)}$ and $\epsilon^{(S)}_{i,c_S} = \nu_{i,c_S}^{(S)} \kappa_{i,c_S}^{(S)}$ where $\kappa_{i,c_T}^{(T)},\kappa_{i,c_S}^{(S)} \in \{1,\sqrt{2}, \sqrt{3}\}$ captures heteroskedasticity and $\rho \in \{0.4, 0.6, 0.8\}$ the strength of the endogeneity. The networks capturing interrelation $\bm{G}_{ST,c_T}$, $\bm{G}_{TS, c_S}$ are sparse where only at maximum only one entry in each row is randomly set to one and zero otherwise.

In each Monte Carlo experiments, $n_{\text{sim}} = 500$ are conducted. The reported performance criteria for each structural parameter or coefficient $\delta \in \bm{\delta} = (\bm{\delta}_T^\prime, \bm{\delta}_S^\prime)$ are the bias, root mean squared error (RMSE) and empirical standard error (ESE) defined as
\begin{align*}
    \text{Bias} &= \frac{1}{n_{\text{sim}}} \sum_{i = 1}^{n_{\text{sim}}} \hat{\delta}_i - \delta \\
    \text{RMSE} &= \sqrt{\frac{1}{n_{\text{sim}}} \sum_{i = 1}^{n_{\text{sim}}} (\hat{\delta}_i - \delta)^2} \\
    \text{ESE} &= \sqrt{\frac{1}{n_{\text{sim}} - 1} \sum_{i = 1}^{n_{\text{sim}}} (\hat{\delta}_i - \bar{\delta})^2}.
\end{align*}
For all proposed performance criteria, lower values are always preferred.

The simulation study is performed in the programming language \textbf{R} \citep{R2025}. Logistic regression is conducted via the \textbf{lme4} package \citep{lme4}. The code for the reproducibility of all results in the Monte Carlo experiments as well as the proposed estimators is available upon request.

\subsection{Results}
The results of the Monte Carlo experiments shown in Table \ref{tab:sim_weak} summarize the performance of the 2SLS and 2SLS-EC estimators in a setting with weak network externalities and endogenous network formation. 

The 2SLS estimator exhibits a systematic downward bias for all structural parameters. The effect is particularly pronounced for the network externalities $\lambda_T$ and $\lambda_S$ whose biases increase substantially as the endogeneity parameter $\rho$ becomes larger. For example, the bias in $\lambda_T$ increases from $-0.132$ at $\rho = 0.4$ to $-0.216$ at $\rho = 0.8$ with an analogous pattern observed for $\lambda_S$. These strong distortions are expected, since these parameters are directly affected by endogenous network formation and thus highly sensitive to self-selection. The cross-network spillover parameters $\lambda_{TS}$ and $\lambda_{ST}$ display smaller mean biases in absolute terms but their estimates show very high variability. The RMSE and ESE values for these parameters are large across all values of $\rho$, indicating that the 2SLS estimator suffers from substantial instability in recovering these cross-equation effects.

The coefficients of exogenous variables are also biased under 2SLS estimator. However, the extent is much less given the relative magnitude of the true coefficients. As $\rho$ increases, both RMSE and ESE increase and in some cases the uncertainty grows sharply (see, for example, for $\gamma_{T,2}$ and $\gamma_{S,2}$). Overall, the results show that 2SLS becomes increasingly unreliable as network endogeneity strengthens and the estimator is particularly noisy for parameters involving interdependence across networks.

In contrast, the 2SLS-EC estimator performs markedly better across all Monte Carlo experiments. The endogeneity correction step almost completely eliminates the downward biases observed under 2SLS. For instance, the bias in $\lambda_T$ at $\rho = 0.8$ decreases from $-0.216$ under 2SLS to only $-0.057$ under 2SLS-EC, and a similar improvement occurs for $\lambda_S$. The biases for the parameters of exogenous variables become negligible, typically well below an absolute value of $0.03$. The 2SLS-EC estimator also yields substantially lower RMSE and ESE for all parameters and all values of $\rho$. This indicates that the estimator is not only centered more accurately around the true parameter values but also much more precise. 

Taken together, the results demonstrate that the 2SLS estimator is generally unreliable in the presence of endogenous network formation, even when network effects are weak. In contrast, the 2SLS-EC estimator consistently removes the bias and improves precision across all structural parameters, confirming that the endogeneity correction procedure is essential for valid inference in this setting.

\begin{table}[!htbp]
\centering
\caption{\label{tab:sim_weak} Monte Carlo experiments for fixed sample sizes $\bar{c}_T = \bar{c}_S = 10$, each of equal size $n_{T,c_T} = n_{S,c_S} = 30$ with weak network effects $\lambda_T = 0.3$, $\lambda_S = 0.2$ and $500$ repetitions. Each parameter block reports bias, root mean squared error (RMSE) (in parentheses), and empirical standard error (ESE) [in brackets] for two-stage least squares (2SLS) and endogeneity-corrected two-stage least squares (2SLS-EC) across $\rho \in \{0.4, 0.6, 0.8\}$.}
\begin{tabular}{lcccccc}
\\[-1.8ex] \hline
\hline \\[-1.8ex]
& \multicolumn{3}{c}{2SLS} & \multicolumn{3}{c}{2SLS-EC} \\
\cmidrule(lr){2-4} \cmidrule(lr){5-7}
& $\rho = 0.4$ & $\rho = 0.6$ & $\rho = 0.8$ & $\rho = 0.4$ & $\rho = 0.6$ & $\rho = 0.8$ \\
\hline \\[-1.8ex]

\multirow{3}{*}{$\lambda_{T} = 0.3$} 
& -0.132 & -0.185 & -0.216 & -0.032 & -0.044 & -0.057 \\
& (0.148) & (0.200) & (0.235) & (0.109) & (0.102) & (0.127) \\
& [0.067] & [0.076] & [0.091] & [0.104] & [0.092] & [0.113] \\

\\[-1.8ex] \hline \\[-1.8ex]

\multirow{3}{*}{$\lambda_{TS} = 0.1$} 
& -0.013 & 0.009 & -0.023 & -0.002 & 0.012 & -0.004 \\
& (0.323) & (0.316) & (0.332) & (0.253) & (0.231) & (0.228) \\
& [0.323] & [0.316] & [0.332] & [0.254] & [0.231] & [0.228] \\

\\[-1.8ex] \hline \\[-1.8ex]

\multirow{3}{*}{$\gamma_{T,1} = 1$} 
& -0.016 & -0.021 & -0.025 & -0.008 & -0.008 & -0.014 \\
& (0.091) & (0.092) & (0.118) & (0.084) & (0.080) & (0.080) \\
& [0.090] & [0.089] & [0.116] & [0.083] & [0.080] & [0.079] \\

\\[-1.8ex] \hline \\[-1.8ex]

\multirow{3}{*}{$\gamma_{T,2} = 0.5$} 
& 0.068 & 0.149 & 0.207 & 0.017 & 0.032 & 0.054 \\
& (0.258) & (0.337) & (0.405) & (0.236) & (0.230) & (0.266) \\
& [0.250] & [0.303] & [0.348] & [0.235] & [0.228] & [0.261] \\

\\[-1.8ex] \hline \\[-1.8ex]

\multirow{3}{*}{$\lambda_{S} = 0.2$} 
& -0.113 & -0.155 & -0.188 & -0.038 & -0.059 & -0.074 \\
& (0.116) & (0.156) & (0.189) & (0.104) & (0.099) & (0.098) \\
& [0.027] & [0.022] & [0.024] & [0.097] & [0.080] & [0.063] \\

\\[-1.8ex] \hline \\[-1.8ex]

\multirow{3}{*}{$\lambda_{ST} = 0.1$} 
& -0.027 & -0.019 & -0.016 & -0.022 & -0.003 & -0.017 \\
& (0.446) & (0.469) & (0.395) & (0.450) & (0.358) & (0.307) \\
& [0.445] & [0.469] & [0.395] & [0.449] & [0.358] & [0.307] \\

\\[-1.8ex] \hline \\[-1.8ex]

\multirow{3}{*}{$\gamma_{S,1} = 1$} 
& -0.049 & -0.065 & -0.082 & -0.015 & -0.023 & -0.034 \\
& (0.094) & (0.103) & (0.110) & (0.095) & (0.093) & (0.083) \\
& [0.080] & [0.080] & [0.074] & [0.093] & [0.090] & [0.076] \\

\\[-1.8ex] \hline \\[-1.8ex]

\multirow{3}{*}{$\gamma_{S,2} = 0.5$} 
& 0.062 & 0.080 & 0.074 & 0.021 & 0.023 & 0.022 \\
& (0.234) & (0.222) & (0.230) & (0.232) & (0.192) & (0.174) \\
& [0.225] & [0.207] & [0.218] & [0.231] & [0.191] & [0.173] \\
\\[-1.8ex] 
\hline \\[-1.8ex]
\end{tabular}
\end{table}

Similarly, in Table \ref{tab:sim_strong}, the results for the Monte Carlo experiments with strong network effects are presented under the assumption that networks are truly endogenous. Overall, the pattern observed is broadly consistent with the Monte Carlo experiments under weak network effects. Thus, magnitude of bias in the 2SLS estimator is generally larger. In particular, the network externality parameters $\lambda_T$ and $\lambda_S$ and the coefficients of exogenous variables $\gamma_{T,1}$, $\gamma_{T,2}$ show notable downward biases as endogeneity $\rho$ increases. Additionally, structural parameter of interrelation $\lambda_{TS}$ and $\lambda_{ST}$ exhibit biases too, although the general direction is varying with endogeneity parameter $\rho$.

The 2SLS-EC estimator effectively mitigates these biases across all parameters. Most mean biases are close to zero, even at the highest level of endogeneity and the spread of the estimates captured via RMSE and ESE is substantially reduced compared to the 2SLS estimator. Small residual biases exist for a few parameters (see, for example, $\lambda_T$ and $\gamma_{T,1}$ at $\rho = 0.8$), but they are minor relative to the corresponding biases in 2SLS, indicating that the endogeneity correction successfully accounts for network endogeneity as in the case of the weak network effects. Thus, the 2SLS-EC estimator remains robust across both weak and strong network effect scenarios, consistently reducing bias and improving estimation precision.

\begin{table}[!htbp]
\centering
\caption{\label{tab:sim_strong} Monte Carlo experiments for fixed sample sizes $\bar{c}_T = \bar{c}_S = 10$, each of equal size $n_{T,c_T} = n_{S,c_S} = 30$ with strong network effects $\lambda_T = 0.6$, $\lambda_S = 0.5$ and $500$ repetitions. Each parameter block reports bias, root mean squared error (RMSE) (in parentheses), and empirical standard error (ESE) [in brackets] for two-stage least squares (2SLS) and endogeneity-corrected two-stage least squares (2SLS-EC) across $\rho \in \{0.4, 0.6, 0.8\}$.}
\begin{tabular}{lcccccc}
\\[-1.8ex] \hline
\hline \\[-1.8ex]
& \multicolumn{3}{c}{2SLS} & \multicolumn{3}{c}{2SLS-EC} \\
\cmidrule(lr){2-4} \cmidrule(lr){5-7}
& $\rho = 0.4$ & $\rho = 0.6$ & $\rho = 0.8$ & $\rho = 0.4$ & $\rho = 0.6$ & $\rho = 0.8$ \\
\hline \\[-1.8ex]

\multirow{3}{*}{$\lambda_{T} = 0.6$} 
& -0.060  & -0.136 & -0.198 & -0.012 & -0.003 & -0.020 \\
& (0.183) & (0.279) & (0.377) & (0.065) & (0.168) & (0.087) \\
& [0.173] & [0.244] & [0.321] & [0.064] & [0.168] & [0.084] \\

\\[-1.8ex] \hline \\[-1.8ex]

\multirow{3}{*}{$\lambda_{TS} = 0.1$} 
& -0.041 & -0.042 & 0.018 & 0.004 & -0.003 & -0.009 \\
& (0.303) & (0.524) & (0.702) & (0.163) & (0.178) & (0.166) \\
& [0.300] & [0.523] & [0.703] & [0.163] & [0.178] & [0.166] \\

\\[-1.8ex] \hline \\[-1.8ex]

\multirow{3}{*}{$\gamma_{T,1} = 1$} 
& -0.057 & -0.113 & -0.153 & -0.006 & -0.007 & -0.021 \\
& (0.285) & (0.378) & (0.459) & (0.121) & (0.174) & (0.134) \\
& [0.279] & [0.361] & [0.433] & [0.121] & [0.174] & [0.132] \\

\\[-1.8ex] \hline \\[-1.8ex]

\multirow{3}{*}{$\gamma_{T,2} = 0.5$} 
& -0.031 & -0.058 & -0.105 & -0.028 & 0.002 & -0.032 \\
& (0.512) & (0.753) & (0.935) & (0.404) & (0.365) & (0.297) \\
& [0.512] & [0.752] & [0.930] & [0.403] & [0.366] & [0.295] \\

\\[-1.8ex] \hline \\[-1.8ex]

\multirow{3}{*}{$\lambda_{S} = 0.5$} 
& -0.042 & -0.079 & -0.112 & -0.004 & -0.004 & -0.008 \\
& (0.123) & (0.257) & (0.219) & (0.048) & (0.062) & (0.056) \\
& [0.116] & [0.245] & [0.188] & [0.048] & [0.062] & [0.055] \\

\\[-1.8ex] \hline \\[-1.8ex]

\multirow{3}{*}{$\lambda_{ST} = 0.1$} 
& 0.011 & 0.003 & -0.050 & -0.008 & -0.014 & 0.001 \\
& (0.299) & (0.598) & (0.726) & (0.133) & (0.211) & (0.205) \\
& [0.300] & [0.598] & [0.725] & [0.133] & [0.210] & [0.206] \\

\\[-1.8ex] \hline \\[-1.8ex]

\multirow{3}{*}{$\gamma_{S,1} = 1$} 
& -0.007 & -0.056 & -0.057 & 0.002 & 0.003 & -0.016 \\
& (0.188) & (0.337) & (0.364) & (0.117) & (0.126) & (0.131) \\
& [0.188] & [0.333] & [0.359] & [0.117] & [0.126] & [0.130] \\

\\[-1.8ex] \hline \\[-1.8ex]

\multirow{3}{*}{$\gamma_{S,2} = 0.5$} 
& -0.006  & 0.007 & -0.038 & -0.003 & 0.007 & 0.007 \\
& (0.476) & (0.670) & (0.833) & (0.283) & (0.309) & (0.303) \\
& [0.476] & [0.670] & [0.833] & [0.283] & [0.309] & [0.304] \\

\\[-1.8ex]
\hline \\[-1.8ex]
\end{tabular}
\end{table}

Thus, if the networks are truly endogenous, then ignoring the network formation process in the construction of the IV matrices leads to the violation of the linear moment conditions. Indeed, the results for the 2SLS estimator exhibit an systematic bias in the estimates of the structural parameters which in turn indicates that the matrices of IV are no longer valid. Additionally, the severity of the consequences of endogenous network formation is varying if weak or strong network effects are considered. In general, the systematic biases can be mitigated via the proposed two-step estimation strategy which essentially relies on replacing the social adjacency matrices by the predicted probabilities of link formation based on the endogenous network formation model. Moreover, the results show proper functionality of the proposed and implemented two-step estimation strategy and indicate that valid sources of IV matrices can be constructed based on the predicted adjacency matrices in the 2SLS-EC estimator. The success of the two-step estimation strategy additionally depends on the considered network effects. Particularly, if the network effects are large and the magnitudes of correlation between unobserved heterogeneity and idiosyncratic random noise are strong, then small leftover systematic biases remain. Furthermore, the presented results are in line with previous research on estimation of structural parameters under endogeneity of social adjacency matrices in single equation models (see, for example, \cite{kelejian2014, lee2021}). The key point of the Monte Carlo experiments is that, when working with real-world data, potential endogenous network formation cannot be safely ignored even if the effect of network externalities is low. Therefore, the proposed two-step estimation strategy has to be utilized to mitigate the consequences of network endogeneity in estimation.

\section{Further empirical results} \label{app:case}
\subsection{Additional information on data}
\subsubsection{Cancer Related Papers} \label{app:cancer}
\begin{figure}[H]
    \centering
    \includegraphics[width=\textwidth]{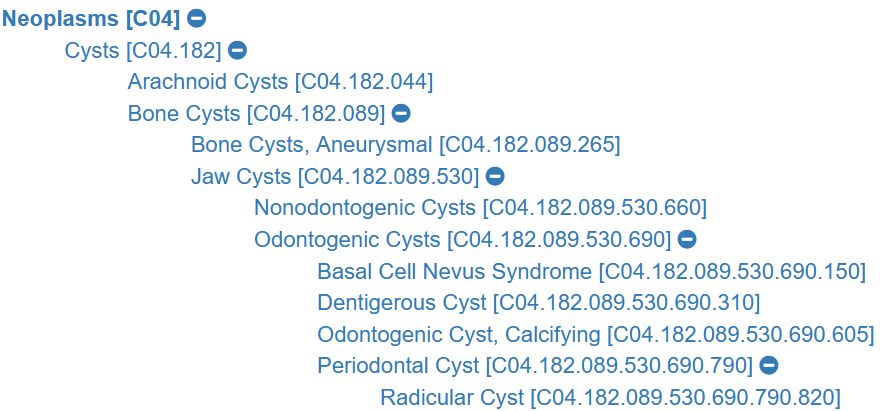} 
    \caption{Example illustration of the MeSH tree structure for Neoplasms child nodes.}
    \label{fig:mesh_tree}
\end{figure}

To identify cancer-relevant publications, we rely on the MeSH terms. Since professional indexers at the National Library of Medicine (NLM) assign MeSH terms to biomedical articles based on standardized protocols, consistency and contextual relevance across the entire database can be ensured. Notably, the authors of these articles do not participate in the assignment of MeSH terms, thereby ensuring an objective indexing process. Additionally, trained professionals oversee this task, reducing the subjectivity that can arise in categorization. Therefore, MeSH terms are an invaluable resource in biomedical research, enabling researchers to navigate and categorize extensive collections of literature efficiently. By organizing articles into a structured hierarchy, MeSH terms facilitate precise and systematic retrieval of information. This capability is crucial for advancing scientific discovery and improving clinical practices. The hierarchical nature of MeSH terms allows for the grouping of related concepts, making it possible to explore both broad themes and specific details within the literature. Cancer-related papers are identified primarily using the child of the MeSH code C04, that is, "Neoplasms," which represents the most comprehensive family of MeSH terms for cancer. This category encompasses a wide range of malignancies, including solid tumors and hematologic cancers, providing a broad foundation for cancer research. Figure \ref{fig:mesh_tree} illustrates the hierarchical structure of MeSH terms related to cancer, demonstrating how specific terms nest under broader categories. This visual representation clarifies the relationships and interconnectedness among MeSH terms, emphasizing their utility and relevance to cancer research.

In addition to MeSH code C04, we consider the advice of actual oncologists to include children of the following MeSH terms, which are relevant to various aspects of cancer biology, treatment, and research:

\begin{itemize}
    \item \textbf{Drug Screening Assays, Antitumor} (E05.337.550.200): These assays are crucial for identifying and developing new antitumor agents by testing their efficacy in inhibiting cancer cell growth.
    \item \textbf{Cancer Vaccines} (D20.215.894.200): Vaccines designed to prevent or treat cancer by stimulating the immune system to target cancer cells.
    \item \textbf{Neoplasms} (C04): A broad category encompassing all types of tumors, both benign and malignant.
    \item \textbf{DNA, Neoplasm} (D13.444.308.425): Refers to the genetic material specific to cancer cells, which can be targeted for diagnosis or therapy.
    \item \textbf{Drug Resistance, Neoplasm} (G07.690.773.984.395): The ability of cancer cells to resist the effects of chemotherapy, posing a significant challenge in cancer treatment.
    \item \textbf{Neoplasm Proteins} (D12.776.624): Proteins specifically associated with tumors, which can serve as biomarkers or therapeutic targets.
    \item \textbf{Biomarkers, Tumor} (D23.101.140): Biological molecules found in blood, other body fluids, or tissues that indicate the presence of cancer.
    \item \textbf{Antigens, Neoplasm} (D23.050.285): Substances produced by tumor cells that can trigger an immune response.
    \item \textbf{Oncogenic Viruses} (B04.613): Viruses that can cause cancer by integrating their genetic material into the host genome.
    \item \textbf{Tumor Cells, Cultured} (A11.251.860): Cancer cells grown in laboratory conditions for research purposes.
    \item \textbf{Neoplasm Proteins} (D12.776.624): Proteins associated with tumors, important for understanding cancer biology and developing treatments.
    \item \textbf{Chemotherapy, Cancer, Regional Perfusion} (E04.292.425): A technique to deliver high doses of chemotherapy directly to the tumor site.
    \item \textbf{Antineoplastic Agents} (D27.505.954.248): Drugs used to treat cancer by inhibiting the growth of malignant cells.
    \item \textbf{Receptors, Tumor Necrosis Factor} (D12.776.543.750.705.852.760): Receptors involved in the signaling pathways that can lead to tumor cell death.
    \item \textbf{Tumor Escape} (G12.900): Mechanisms by which cancer cells evade the immune system.
    \item \textbf{Neoplastic Stem Cells} (A11.872.650): Stem cells within tumors that have the ability to self-renew and drive cancer progression.
    \item \textbf{Carcinogens} (D27.888.569.100): Substances capable of causing cancer in living tissue.
    \item \textbf{Gammaretrovirus} (B04.820.650.375): A type of virus that can insert its genetic material into the host genome, potentially causing cancer.
    \item \textbf{Antibodies, Neoplasm} (D12.776.377.715.548.114.240): Antibodies used to target and neutralize cancer cells.
    \item \textbf{Receptors, Immunologic} (D12.776.543.750.705): Receptors on immune cells that can be manipulated to enhance the immune response against cancer.
    \item \textbf{Tumor Necrosis Factors} (D23.529.374.750): Proteins involved in the destruction of cancer cells.
    \item \textbf{Biomarkers, Tumor} (D23.101.140): Indicators used to detect cancer or monitor its progression.
    \item \textbf{Radiotherapy} (E02.815): The use of high-energy radiation to kill or shrink cancer cells.
\end{itemize}

\bibliography{sn-bibliography}

\end{document}